\pdfoutput=1
\documentclass[
    affil-it, 
    auth-lg, 
    british, 
    compression, 
    contents, 
    custom-packages={mleftright}, 
    references={bigone}, 
]{lib/preprint}

\hypersetup{
    pdftitle = {Higher‐order analogues of colour–kinematics duality in first‐order Yang–Mills Feynman diagrams},
    pdfauthor = {Leron Borsten, Dimitri Kanakaris, Hyungrok Kim},
    pdfkeywords = {double copy, color–kinematics duality, colour–kinematics duality, BV◼-algebras, Batalin–Vilkovisky algebras},
}

\tikzset{
    gluon/.style={
        decorate, 
        draw=black,
        decoration={
            snake,
            post=lineto,
            post length=0pt,
            segment length=4,
            amplitude=0.9
        }
    }
}

\usepackage{microtype}
\newcommand{\makecommand}[3]{%
	\foreach \i in #3 {%
    	\expandafter\xdef\csname #1\i\endcsname{{\noexpand#2{\unexpanded\expandafter{\i}}}}%
  	}%
}
\makecommand{fr}{\mathfrak}{{g,A,F,I,L}}
\makecommand{sf}{\mathsf}{{a,b,c,d,e,f,g,m,n,r,s,w,P,Q,V}}
\makecommand{sc}{\mathscr}{{A}}

\DeclareMathOperator{\id}{\mathrm{id}}
\DeclareMathOperator{\im}{im}

\DeclareMathOperator{\intprod}{\mathbin{\raisebox{\depth}{\scalebox{1}[-1]{$\lnot$}}}}
\newcommand{\eand}{{~~~\mbox{and}~~~}}
\makecommand{sf}{\mathsf}{{Aut}}
\newcommand{\gh}{{\text{gh}}}
\newcommand{\BV}{{\text{BV}}}
\newcommand{\BRST}{{\text{BRST}}}
\newcommand{\extd}{\mathrm{d}}
\newcommand{\Lie}{\operatorname{Lie}}
\newcommand{\ym}{\textsc{ym}}
\newcommand{\hMC}{{\text{hMC}}}
\newcommand{\braket}[2]{\left\langle #1\,,#2\right\rangle}
\newcommand{\bbR}{\mathbb{R}}
\newcommand{\Order}{\operatorname{Order}}
\newcommand{\vol}{\operatorname{vol}}
\newcommand{\dg}{\dagger}
\newcommand{\dR}{\mathrm{dR}}
\newcommand{\tT}{\mathrm{T}}
\newcommand{\cC}{\mathcal{C}}
\newcommand{\bbZ}{\mathbb{Z}}

\newcommand{\vsum}[2]{
\begin{gathered}
	#1
	\\[-0.5em]
	\oplus
	\\[-0.5em]
	#2
\end{gathered}
}
\newcommand{\vsumt}[1]{
\begin{gathered}
	#1
	\\[-0.5em]
	\phantom\oplus
	\\[-0.5em]
	\phantom{\Omega^d}
\end{gathered}
}
\newcommand{\vsumb}[1]{
\begin{gathered}
	\phantom{\Omega^d}
	\\[-0.5em]
	\phantom\oplus
	\\[-0.5em]
	#1
\end{gathered}
}

\newcommand{\pder}[2]{\frac{\partial #1}{\partial #2}}
\newcommand{\tpder}[2]{\tfrac{\partial #1}{\partial #2}}

\begin{document}
    \date{\today}
    \preprint{}
    \email{l.borsten@herts.ac.uk, d.kanakaris-decavel@herts.ac.uk, h.kim2@herts.ac.uk}
    \title{Higher-order analogues of colour--kinematics duality in first-order Yang--Mills Feynman diagrams} 
    \author[a]{Leron~Borsten\,\orcidlink{0000-0001-9008-7725}\,}
    \author[a]{Dimitri~Kanakaris\,\orcidlink{0009-0001-7716-851X}\,}
    \author[a]{Hyungrok~Kim\,\orcidlink{0000-0001-7909-4510}\,}
    \affil[a]{Centre for Theoretical Physics Research, Department of Physics, Astronomy, and Mathematics, University of Hertfordshire, Hatfield AL10 9AB, United Kingdom}
    \abstract{
    Colour--kinematics duality is, even at tree level, not manifest from the standard Lagrangian of Yang--Mills theory in that the usual Feynman-diagram expansion does not follow the kinematic Jacobi identities.
    For a Lagrangian manifesting colour--kinematics duality, the kinematic Jacobi identities of the Feynman-diagram expansion follow from the existence of a second-order differential operator \(\sfb\) acting on the algebra of colour-stripped fields that forms part of the data of a BV\(^\Box\)-algebra.
    We argue that the existence of differential operators of order greater than two implies weaker but nontrivial fragments of kinematic Jacobi identities. We further show that a superspace formulation of the first-order Yang--Mills action in the Batalin--Vilkovisky formalism admits a differential operator of order six in every spacetime dimension.
    Therefore, the Feynman-diagram expansion of tree and loop scattering amplitudes of the first-order formulation of Yang--Mills theory
    automatically enjoy a weak form of colour--kinematics duality.
    }
    \acknowledgements{The authors thank Branislav Jurčo\,\orcidlink{0000-0001-7782-2326}, Christian Saemann\,\orcidlink{0000-0002-5273-3359}, and Martin Wolf\,\orcidlink{0009-0002-8192-3124} for helpful discussions.}
    \datalicencemanagement{No additional research data beyond the data presented and cited in this work are needed to validate the research findings in this work.}
\begin{body}
    
\section{Introduction and summary}

\paragraph{Background.}
Colour--kinematics (CK) duality \cite{Bern:2008qj,Bern:2010ue,Bern:2010yg} (reviewed in~\cite{Carrasco:2015iwa,Bern:2019prr,Borsten:2020bgv,White:2021gvv,Adamo:2022dcm,Bern:2022wqg, Lescano:2026xlb})
is a hidden symmetry of a wide class of gauge theories, including for tree-level pure Yang--Mills theory in an arbitrary number of spacetime dimensions,
in which scattering amplitudes admit a nontrivial reorganisation in which the momenta factors in the numerators, the so-called \emph{kinematic numerators}, admit antisymmetry and Jacobi identities, similar to the colour factors in the numerators. This suggests the existence of a kinematic Lie algebra, which is indeed known explicitly  in certain cases \cite{Monteiro:2011pc,Monteiro:2013rya,Bjerrum-Bohr:2012kaa,Cheung:2016prv, Ben-Shahar:2021zww,Borsten:2022vtg,Borsten:2023reb,Bonezzi:2023pox}. 
CK-dual gauge theories have a remarkable connection to gravitational theories, a phenomenon known as the \emph{double copy} \cite{Bern:2008qj,Bern:2010ue,Bern:2010yg}: if the colour factors are replaced with another copy of the kinematic numerators in the in the CK-dual representation of the gauge-theory amplitudes, one obtains the scattering amplitudes of a gravitational theory. Thus one would like to known what theories exhibit colour--kinematics duality.

This is not generally a simple question. Even at tree level, the colour--kinematics duality of  the on-shell amplitudes is not directly visible from the ordinary action of Yang--Mills theory: in order to manifest it, one has to add an infinite series of terms that formally vanish due to the colour Jacobi identities but, when colour-stripped, lead to nontrivial Feynman diagrams that correct the Feynman-diagram expansion of the usual Yang--Mills action in a CK-dual fashion  \cite{Bern:2010yg,Tolotti:2013caa}. The on-shell tree-level amplitudes constructed directly from the Feynman rules then obey colour--kinematics duality automatically. In fact, for any theory  whose  tree-level   S-matrix admits a CK-dual representation, there exists a purely cubic Batalin--Vilkovisky  action whose off-shell Feynman rules for all states (physical, auxiliary, gauge and ghost) satisfy colour--kinematics duality \cite{Borsten:2020zgj, Borsten:2021hua,Borsten:2021gyl}. This action will typically require an infinite tower of auxiliary fields\footnote{A notable, and conceptually instructive, exception is Chern-Simons theory \cite{Ben-Shahar:2021zww}.}.  Since this is a statement regarding the off-shell Feynman rules for all states,  the loop-level kinematic numerators automatically obey colour--kinematics duality, although it may be  anomalous\footnote{In the sense that any unitarity respecting regulation scheme  will  break colour--kinematics duality.} \cite{Borsten:2021gyl}.  

The relationship between colour--kinematics duality as a hidden property of the on-shell scattering amplitudes and as a conventional (albeit generically infinite dimensional and potentially anomalous) symmetry of the Batalin--Vilkovisky  action is naturally captured through the theory of  homotopy algebras \cite{Borsten:2020zgj, Borsten:2021hua,Borsten:2021gyl, Escudero:2022zdz,Borsten:2022vtg,Bonezzi:2022bse,Bonezzi:2023lkx,Borsten:2023paw,Szabo:2023cmv,Bonezzi:2023pox,Bonezzi:2024dlv,Borsten:2024cfx,Ben-Shahar:2025dci}. Every perturbative Lagrangian quantum field theory is equivalent, in a precise sense, to an equivalence class of homotopy algebras. A  standard action and its colour--kinematics duality manifesting cousin are then merely  representatives of the same equivalence class;  colour--kinematics duality was only ever hidden up to homotopy.

More precisely, from the homotopy algebra perspective colour--kinematics duality can be formulated in the language of (homotopy) BV\(^\Box\)-algebras \cite{Reiterer:2019dys,Borsten:2021gyl, Borsten:2022vtg,Bonezzi:2022bse,Bonezzi:2023lkx,Borsten:2023paw,Bonezzi:2023pox,Bonezzi:2024dlv,Borsten:2024cfx,Ben-Shahar:2025dci,Diaz-Jaramillo:2025gxw, Bonezzi:2026qxf}. In particular, \cite{Reiterer:2019dys} demonstrated that  Yang--Mills theory (specifically in four dimensions) carries an intricate homotopy BV\(^\Box_\infty\)-algebra structure, defined therein\footnote{Recent work \cite{medinamardones2025operadiccalculushighercolourkinematics} gave  a first-principles definition of  the relevant (homotopy) operad, named cBV$_\infty$ therein, rigorously establishing the homotopical machinery of deformation
theory, homotopy transfer, rectification,  minimal models, and tensor products for cBV$_\infty$/BV\(^\Box_\infty\)-algebras. The tensor product was shown to agree with the  product of metric BV\(^\Box\)-algebras  introduced in \cite{Borsten:2023ned} for the double copy.}, which implies the tree-level colour--kinematics duality of the S-matrix. It was subsequently shown that, more generally, a theory with purely cubic interactions (using auxiliary fields if necessary) has a colour--kinematics duality manifesting Batalin-Vilkovisky action if and only if it possesses a BV\(^\Box\)-algebra structure \cite{Borsten:2021gyl, Borsten:2022vtg, Borsten:2023ned}.  The key non-trivial datum of a  BV\(^\Box\)-algebra is a second-order differential operator $\sfb$ (the notation reflects its analogous role to the $b$-ghost in string theory)  that plays nice with the free equations of motion and forms the numerator of the propagator (see \cref{ckdual}).  Colour--kinematics duality then manifests itself as a  kinematic Lie algebra generated by  a derived bracket with respect to $\sfb$. This picture was  developed in \cite{Borsten:2023ned}, where it was also noted that if $\sfb$ is a higher-order differential operator, the kinematic Lie algebra is weakened to a kinematic homotopy Lie algebra ($L_\infty$-algebra) realised through a standard construction, the Koszul hierarchy \cite{koszul1985crochet,Akman:1995tm,Bering:1996kw,Markl:2013pca}. The equivalent codifferential picture  was used in \cite{Ben-Shahar:2024dju} to show  that  four-dimensional $BF$-theory and two-dimensional Yang--Mills theory in the standard first-order formalism both admit a second-order $\sfb$, and  the possibility of higher-order differentials for Yang--Mills theory in other dimensions was also noted. 
        
\paragraph{Summary of results.}
Since ordinary colour--kinematics duality of the Feynman-diagram expansion is indicated by the presence of a second-order differential operator $\sfb$ on the algebra of colour-stripped fields, it is natural to consider weakened versions of colour--kinematics duality implied by the existence of an \(n\)\textsuperscript{th}-order differential operator for \(n>2\). To the best of our knowledge, there are no known explicit examples.  

We explain that for a generic quantum field theory with purely cubic interactions\footnote{All theories with a BV action may be put in purely cubic form through auxiliary fields.} in the BV formalism, the existence of a differential operator that squares to zero (i.e.\ $\sfb^2=0$) implies the existence of a kinematic $L_\infty$-algebra given by the Koszul hierarchy corresponding to $\sfb$ and the colour-stripped three-point vertex. Specifically, the diagrams built from the three-point vertex and single insertions of $\sfb$ (which are related to, but identical to, the kinematic numerators) obey the order-$r$ colour--kinematics duality relation exactly, as opposed to holding up to a higher chain homotopy.

These general principles are explicitly demonstrated for pure (i.e.\ non-supersymmetric and without matter)  Yang--Mills theory. We construct a superspace geometric formulation of the first-order action of pure Yang--Mills theory in any number of spacetime dimensions, and show that it naturally admits a differential operator \(\tilde\sfb\) of order six. It does not square to zero, but this can be repaired by the addition of a term that is exact with respect to the Batalin--Vilkovisky differential; the repaired operator \(\sfb\) is still of order six in any number of spacetime dimensions.

The existence of this operator then implies that, for the gauge choice given by \(\sfb\), the Feynman-diagrammatic expansion of Yang--Mills theory, without the addition of any vanishing terms similar to \cite{Tolotti:2013caa, Borsten:2020zgj, Borsten:2021hua, Borsten:2021gyl}, already obey a subset of the kinematic Jacobi identities; in particular, they hold automatically at the loop level as well.

\paragraph{Future directions.}
    In this paper, we confine ourselves to a discussion of pure Yang--Mills theory. A natural next step would be to discuss whether the partial colour--kinematics duality observed here persists or is possibly enhanced in the presence of matter or supersymmetry.
    
    Our results naturally raise the question of whether the weaker fragment of colour--kinematics duality defined here suffices for some version of the double copy. While intriguing, we do not discuss this question in this paper.

\paragraph{Organisation of this paper.}
After a brief summary of the Batalin--Vilkovisky formalism, (on- and off-shell) colour--kinematics duality and its homotopy-algebraic formulation in \cref{sec:review} (for more details, the reader may consult e.g.~\cite{Borsten:2023ned}), we introduce a notion of higher-order colour--kinematics duality via higher-order differentials and the Koszul hierarchy in \cref{sec:koszul}. We explain some of the generic consequences of a finite order $\sfb$ and that the  Koszul hierarchy forms an $L_\infty$-algebra. We then present  the diagrammatic identities that such an operator implies for certain off-shell tree and loop diagrams built from the colour-stripped three-point vertex and propagator numerator.  We bring these construction together in \cref{sec:main}, 
where we explain how the usual action of pure Yang--Mills theory arises from a certain modified algebra of differential forms and, using this perspective, write down a sixth-order differential operator \(\sfb\) that manifests sixth-order colour--kinematics duality.
Detailed conventions used in the paper are given in \ref{sec:Conventions}.
For the sake of presentability, detailed computations are relegated to \ref{sec:Computations}.

\section{The Batalin--Vilkovisky formalism and colour--kinematics duality}\label{sec:review}
\subsection{The Batalin--Vilkovisky formalism and homotopy algebras}

The Batalin--Vilkovisky (BV) formalism \cite{Batalin:1977pb,Batalin:1981jr,Batalin:1983ggl,Batalin:1984ss,Batalin:1985qj} (reviewed in \cite{Henneaux:1992,Qiu:2011qr}) provides an elegant formulation of quantum field theories with gauge symmetries. For every physical field (of ghost number zero) and Faddeev--Popov ghost (of positive ghost number), one introduces a corresponding \emph{antifield} (not to be confused with antiparticles); the antifield \(\phi^+\) of a field \(\phi\) of ghost number \(k\) has ghost number \(-1-k\).
The BV formalism dovetails with the theory of homotopy algebras: the space of perturbative fields in the BV formalism
form a cyclic \(L_\infty\)-algebra, and the BV action can be recovered as the homotopy Maurer--Cartan action of this cyclic \(L_\infty\)-algebra \cite{Zwiebach:1992ie, Jurco:2018sby}; a field or antifield with ghost number \(k\) corresponds to a component of the \(L_\infty\)-algebra in degree \(1-k\). For instance, the Batalin--Vilkovisky formulation of pure Yang--Mills theory involves fields \(c\) (Faddeev--Popov ghost) and \(A\) (gauge boson) and antifields \(A^+\) and \(c^+\); these sit in an \(L_\infty\)-algebra in degrees \(0, 1, 2, 3\) respectively, corresponding to ghost numbers \(1, 0, -1, -2\) respectively. Note that every perturbative Lagrangian quantum field theory corresponds to a cyclic $L_{\infty}$-algebra via the BV formalism. We briefly mention the essential ingredient here. For a detailed review in the context of field theories, see \cite{Jurco:2018sby}.

Let us denote the antisymmetric Lie bracket $[-,-]\colon \frg^{\otimes 2}\to\frg$ of a conventional Lie algebra by $\mu_2(-,-)$. An \(L_\infty\)-algebra is then the homotopy generalisation of a Lie algebra to a \emph{graded} vector  space with graded antisymmetric brackets of arbitrary arity 
$\mu_k\colon \frL^{\otimes k}\to\frL$. These brackets satisfy homotopy Jacobi relations:
\begin{equation}
\sum_{\substack{j+k=i \\ \sigma \in \operatorname{Sh}(j ; i)}} \chi\left(\sigma ; \phi_1, \ldots, \phi_i\right)(-1)^k \mu_{k+1}\left(\mu_j\left(\phi_{\sigma(1)}, \ldots, \phi_{\sigma(j)}\right), \phi_{\sigma(j+1)}, \ldots, \phi_{\sigma(i)}\right)=0
\end{equation}
for all $i \in \mathbb{N}$ and $\phi_1, \ldots, \phi_i \in \frL$, where $\chi\left(\sigma ; \phi_1, \ldots, \phi_i\right)$ is the graded-antisymmetric Koszul sign and the  sum is over all shuffles $\sigma \in \operatorname{Sh}(j ; i)$.

The first few relations are instructive. For $i=1$, the relation reads $\mu_1(\mu_1(\phi))=0$, so $\mu_1$ is a differential. For $i=2$, it states that $\mu_1$ is a graded derivation of the bracket $\mu_2$. For $i=3$, we have
\begin{equation}
\begin{split}
&\mu_2\big(\mu_2(\phi_1,\phi_2),\phi_3\big)+(-1)^{|\phi_1|(|\phi_2|+|\phi_3|)}\mu_2\big(\mu_2(\phi_2,\phi_3),\phi_1\big) \\
&\quad+(-1)^{|\phi_3|(|\phi_1|+|\phi_2|)}\mu_2\big(\mu_2(\phi_3,\phi_1),\phi_2\big) \\
&=-\Big(\mu_1\mu_3(\phi_1,\phi_2,\phi_3)+\mu_3(\mu_1 \phi_1,\phi_2,\phi_3) \\
&\quad+(-1)^{|\phi_1|}\mu_3(\phi_1,\mu_1 \phi_2,\phi_3)+(-1)^{|\phi_1|+|\phi_2|}\mu_3(\phi_1,\phi_2,\mu_1 \phi_3)\Big).
\end{split}
\end{equation}
The left-hand side is the graded Jacobiator of $\mu_2$; the right-hand side shows that it vanishes only up to $\mu_1$-exact terms governed by $\mu_3$.  We see that, if all higher-arity brackets vanish except for the unary and binary brackets so that the Jacobiator vanishes identically, then an \(L_\infty\)-algebra reduces to a differential graded (dg) Lie algebra.

In the context of BV field theories, $\mu_1$ determines the free equations of motion, while the higher brackets $\mu_k$ encode the $(k+1)$-point interaction terms. However, any \(L_\infty\)-algebra can be converted into a dg Lie algebra that contains equivalent information (technically, any \(L_\infty\)-algebra is quasi-isomorphic to a dg Lie algebra); physically, this corresponds to the introduction of auxiliary fields that reduce $(k+1)$-point interactions  down to purely  to cubic ones. 

Finally, the Feynman-diagram expansion of the tree-level scattering amplitudes\footnote{Loops are including by passing to the corresponding loop \(L_\infty\)-algebra, see  \cite{Macrelli:2019afx,Jurco:2019yfd} and references therein.} of a field theory corresponds to the \emph{homotopy transfer} of the \(L_\infty\)-algebra structure onto its $\mu_1$-cohomology, which physically corresponds to the asymptotic on-shell states \cite{Macrelli:2019afx,Jurco:2019yfd,Saemann:2020oyz,Maunder:2024xeq}. This yields a physically equivalent `minimal model'  \(L_\infty\)-algebra, whose homotopy Maurer Cartan action is the generating function of the tree-level S-matrix.

\subsection{Colour--kinematics duality}\label{ckdual}

\paragraph{On-shell colour--kinematics duality.} Let us first recall the  colour--kinematics duality of scattering amplitudes. An $n$-point, $L$-loop scattering amplitude integrand $\scA_{n,L}$ can be parameterised as 
        \begin{equation}\label{eq:CK_amplitudes_parameterization}
            \scA_{n,L}\ \sim\ \sum_{\gamma\in\Gamma_{n,L}}\int_{\text{loops}}\frac{\sfc_\gamma\sfn_\gamma}{|\sfAut(\gamma)|d_\gamma}~,
        \end{equation}
        where $\Gamma_{n,L}$ is the set of $n$-point, $L$-loop cubic diagrams; $\sfc_\gamma$ is the {colour numerator}, that is, the contribution to the diagram $\gamma$ due to the metric and the structure constants of the gauge Lie algebra; $d_\gamma$ is the product of the denominators of the propagators for $\gamma$; $|\sfAut(\gamma)|$ is the symmetry factor of the diagram $\gamma$, i.e.~the order of its automorphism group; and $\sfn_\gamma$ is the {kinematic numerator} containing the remaining contributions of $\gamma$ to $\scA_{n,L}$ and a function of kinematic data only. 
        
        The antisymmetry and Jacobi identity of the Lie algebra structure constants imply that  certain sums of colour numerators vanish, i.e.
        \begin{equation}\label{eq:CKcolor}
            \sfc_{\gamma_{a1}}+\sfc_{\gamma_{a2}}\ =\ 0
            \eand
            \sfc_{\gamma_{J1}}+\sfc_{\gamma_{J2}}+\sfc_{\gamma_{J3}}\ =\ 0
        \end{equation}
        for certain pairs $(\gamma_{a1},\gamma_{a2})$ and triples $(\gamma_{J1},\gamma_{J2},\gamma_{J3})$. Specifically, $\sfc_{\gamma_{J1}}+\sfc_{\gamma_{J2}}+\sfc_{\gamma_{J3}}\ =\ 0$ holds for any triple of diagrams $(\gamma_{J1},\gamma_{J2},\gamma_{J3})$ that only differ in a common four-point subdiagram, with  the $s$-, $t$- and $u$-channel diagrams inserted:
         \begin{equation}
            \gamma_{J1}= \begin{tikzpicture}[
                scale=1,
                every node/.style={scale=1},
                baseline={([yshift=-.5ex]current bounding box.center)}
                ]
                \matrix (m) [
                matrix of nodes,
                ampersand replacement=\&,
                column sep=0.13cm,
                row sep=0.13cm
                ]{
                    {} \& {}\& {} \& {} \& {}
                    \\
                    {} \& {} \& {}\& {} \& {}
                    \\
                    {} \& {} \& {} \& {} \& {}
                    \\
                    {} \& {} \& {}\& {} \& {}
                    \\
                    {} \& {} \& {} \& {} \& {}
                    \\
                };
                \draw [gluon] (m-1-1) -- (m-3-2.center);
                \draw [gluon] (m-5-1) -- (m-3-2.center);
                \draw [gluon] (m-3-2.center) -| node[near start,below] {} (m-3-4.center);
                \draw [gluon] (m-1-5) -- (m-3-4.center);
                \draw [gluon] (m-5-5) -- (m-3-4.center);
                \node [circle,draw,dashed,minimum size=1.9cm] (c) at (0,0){};
                \foreach \x in {(m-3-2), (m-3-4)}{
                    \fill \x circle[radius=2pt];
                }
            \end{tikzpicture}
            ~~~~~~
          \gamma_{J2}=  \begin{tikzpicture}[
                scale=1,
                every node/.style={scale=1},
                baseline={([yshift=-.5ex]current bounding box.center)}
                ]
                \matrix (m) [
                matrix of nodes,
                ampersand replacement=\&,
                column sep=0.13cm,
                row sep=0.13cm
                ]{
                    {} \& {} \& {} \& {} \& {}
                    \\
                    {} \& {} \& {} \& {} \& {}
                    \\
                    {} \& {} \& {} \& {} \& {}
                    \\
                    {} \& {} \& {} \& {} \& {}
                    \\
                    {} \& {} \& {} \& {} \& {}
                    \\
                };
                \draw [gluon] (m-1-1) -- (m-2-3.center);
                \draw [gluon] (m-5-1) -- (m-4-3.center);
                \draw [gluon] (m-2-3.center) -| node[near end,left] {} (m-4-3.center);
                \draw [gluon] (m-1-5) -- (m-2-3.center);
                \draw [gluon] (m-5-5) -- (m-4-3.center);
                \node [circle,draw,dashed,minimum size=1.9cm] (c) at (0,0){};
                \foreach \x in {(m-2-3), (m-4-3)}{
                    \fill \x circle[radius=2pt];
                }
            \end{tikzpicture}
            ~~~~~~
            \gamma_{J3}= \begin{tikzpicture}[
                scale=1,
                every node/.style={scale=1},
                baseline={([yshift=-.5ex]current bounding box.center)}
                ]
                \matrix (m) [
                matrix of nodes,
                ampersand replacement=\&,
                column sep=0.13cm,
                row sep=0.13cm
                ]{
                    {} \& {}\& {} \& {} \& {}
                    \\
                    {} \& {} \& {}\& {} \& {}
                    \\
                    {} \& {} \& {}\& {} \& {}
                    \\
                    {} \& {} \& {}\& {} \& {}
                    \\
                    {} \& {}\& {} \& {} \& {}
                    \\
                };
                \draw [gluon] (m-1-1) -- (m-3-4.center);
                \draw [gluon] (m-5-1) -- (m-3-2.center);
                \draw [gluon] (m-3-2.center) -| node[near start,below] {} (m-3-4.center);
                \draw [gluon] (m-1-5) -- (m-3-2.center);
                \draw [gluon] (m-5-5) -- (m-3-4.center);
                \node [circle,draw,dashed,minimum size=1.9cm] (c) at (0,0){};
                \foreach \x in {(m-3-2),(m-3-4)}{
                    \fill \x circle[radius=2pt];
                }
            \end{tikzpicture}
        \end{equation}
A theory is said to be colour--kinematics-dual if the same relations hold for the corresponding kinematic numerators:
        \begin{equation}\label{eq:CKkinematic}
            \sfn_{\gamma_{a1}}+\sfn_{\gamma_{a2}}\ =\ 0
            \eand
            \sfn_{\gamma_{J1}}+\sfn_{\gamma_{J2}}+\sfn_{\gamma_{J3}}\ =\ 0~.
        \end{equation}
More precisely, a theory is colour--kinematics-dual if there exists some representation of the amplitudes such that these conditions hold. 

Indeed, even if a theory is colour--kinematics-dual, the kinematic numerators generated by the Feynman rules will typically not satisfy the required Jacobi identities. However, if a theory has tree-level colour--kinematics duality, then there exists an equivalent BV action whose Feynman vertices manifest colour--kinematics duality. This lifts colour--kinematics duality to a symmetry of the action, which may, however, be anomalous. This may be regarded as  \emph{full off-shell colour--kinematics duality} in the sense that that the Feynman vertices  for all off-shell fields, including ghosts and auxiliary fields, manifest colour--kinematics duality identically. This means loop diagrams will also be CK-dual, but this may be broken by counter terms required for unitary. See \cite{Borsten:2021gyl} for full details.

\paragraph{Off-shell colour--kinematics duality.}  Let us now  summarise full off-shell colour--kinematics duality from the homotopy-algebraic perspective.  For more details, see for example \cite{Borsten:2020zgj,Borsten:2021hua, Borsten:2021gyl,Borsten:2022vtg,Borsten:2023ned}.

Suppose that we have a perturbative quantum field theory with only cubic interactions. The corresponding cyclic $L_\infty$-algebra $\frL$ is a strict dg Lie algebra with inner product (cyclic structure) $\langle -, - \rangle$, differential $\mu_1$ and Lie bracket $\mu_2$. Note that $\mu_1$ and $\mu_2$ encode the kinetic and cubic  interactions terms in the BV action respectively, which are of the form
\begin{equation}
\langle \phi_1, \mu_1(\phi_2) \rangle, \qquad \langle \phi_1, \mu_2(\phi_2, \phi_3)\rangle.
\end{equation}
See \cite{Jurco:2018sby} and \cref{ssec:Colour-stripping first order Yang-Mills theory} for numerous explicit examples including Yang--Mills theory.

Assume that our theory has a colour Lie algebra $\frg$ so that the corresponding $L_\infty$-algebra  factors,
\begin{equation}
    \mathfrak L = \frA\otimes\frg,
\end{equation}
where \(\frA\) is the colour-stripped dg commutative algebra with differential $\sfd$. Since $\frg$ is a conventional Lie algebra with bracket $[-,-]$, $\mu_1=\sfd\otimes \id_\frg$ and $\mu_2 = \sfm_2(-,-)\otimes [-,-]$, where   $\sfm_2$ is the binary graded commutative product on $\frA$. Note that $\sfm_2$ is nothing but the colour-stripped Feynman vertex for all fields\footnote{More precisely, $\sfm_2$ is a map from pairs of fields to  antifields, which are then mapped back to fields by the propagator (or, equivalently, the contracting homotopy of the homological perturbation lemma that provides the quasi-isomorphism to the minimal model). So it is $\sfb\circ \sfm_2$ that maps pairs of fields to fields. The colour-stripped three-point vertex as usually given in textbooks is actually $\langle -, \sfm_2(-,-)\rangle$ and it is the cyclicity of $\langle -, -\rangle$ with respect to $\sfb$ that ensures these pictures are equivalent.}. For notational convenience, we will often write $ \sfm_2(x,y)=xy$, where $x,y,z,\ldots\in \frA$ are our generic colour-stripped fields. 

Now, if  there exists a degree $-1$ operator $\sfb\colon \frA\to \frA$ such that 
\begin{equation}\label{hodge}
[\sfb, \sfd] = \Box,
\end{equation}
where $\Box$ is the d'Alembertian, and 
\begin{equation}
\begin{split}\label{2ndorder}
{\sfb}(xyz) &= {\sfb}(xy)\,z + (-1)^{|y||z|}\,{\sfb}(xz)\,y + (-1)^{|x|(|y|+|z|)}\,{\sfb}(yz)\,x \\
&~~~- ({\sfb}x)\,yz - (-1)^{|x|}\,x\,({\sfb}y)\,z - (-1)^{|x|+|y|}\,xy\,({\sfb}z)
\end{split}
\end{equation}
then theory has an off-shell kinematic Lie algebra defined on the full set of BV fields, with (shifted) Lie bracket given by the derived bracket
\begin{equation}
[x,y]_{\text{kin}}\coloneq {\sfb}(xy) - ({\sfb}x)\,y - (-1)^{|x|}\,x\,({\sfb}y). 
\end{equation}
This follows from a choice of gauge such  that the propagator is \(\sfb/\Box\) and, thus, \(\sfb\) appears in the kinematic numerator. A dg commutative algebra with such a \(\sfb\) operator is called a (strict) BV\(^\Box\)-algebra; see \cite{Borsten:2023ned} for more details.

If we choose the gauge $\sfb(x)=0$, then $[x,y]_{\text{kin}} = {\sfb}(xy)$ and  ${\sfb}\circ \sfm_2$ is the vertex appearing in the kinematic numerators of the currents given by the Feynman rules\footnote{A priori, $\sfb$ is the numerator of the propagator, but we can equivalently attach it to each vertex in this case.}. Since $[x,y]_{\text{kin}}$  obeys the Jacobi identity for all off-shell fields (including ghosts and auxiliary fields), the current numerators built from it automatically satisfy colour--kinematics duality. If the BV\(^\Box\)-algebra is compatible with the inner product, as described in  \cite{Borsten:2023ned}, then the kinematic numerators for the amplitudes   (off-shell and for all external states, including ghosts) are also built from ${\sfb}\circ \sfm_2$, but with $\sfb$ stripped off the out edge, and will satisfy colour--kinematics duality.

The condition \eqref{2ndorder} can be naturally relaxed; this leads to weaker versions of colour--kinematics duality in which only a subset of the kinematic Jacobi identities is satisfied. We turn to these notions now.

\section{The Koszul hierarchy and colour--kinematics duality}\label{sec:koszul}

A given representation of a field theory in terms of a classical BV action (equivalently, cyclic $L_\infty$-algebra) with purely cubic interactions may not exhibit a strict BV${}^\Box$ algebra structure and thus have no associated kinematic Lie algebra. The corresponding Feynman diagrams will not exhibit colour--kinematics duality. However, there always exists a higher-order notion of colour--kinematics duality. The numerators of the Feynman diagrams will obey higher-order Jacobi-like relations \cite{Borsten:2023ned}. This follows from the Koszul hierarchy associated with higher-order differential operators, as we now explain.

\subsection{Higher-order differential operators and the Koszul hierarchy}
\label{ssec:Higher-order differential operators and the Koszul hierarchy}
Let \(\frA\) be a graded-commutative associative algebra. In direct analogy to differential operators on functions, there is a natural  notion of  \emph{higher-order differential operators} on $\frA$ \cite{koszul1985crochet,Akman:1995tm}, whose definition we now recall (see \cite{Borsten:2023ned} for more details).

\paragraph{Higher-order differentials.} A  \emph{first-order differential operator} \(D\colon\frA\to\frA\) with grading \(|D|\)  satisfies the Leibniz identity. Equivalently, the obstruction to the Leibniz identity
\begin{equation}\label{eq:leibniz identity}
    \Phi^2_D(x,y) \coloneqq D(xy) - (Dx)y - (-1)^{|D||x|}x(Dy)
\end{equation}
vanishes, \(\Phi^2_D(x,y)=0\) for all homogeneous elements \(x,y\in\frA\), so  that $D$ is a graded derivation.

A \emph{second-order differential operator} \(D\colon\frA\to\frA\) satisfies 
\begin{equation}
    \Phi^3_D(x,y,z) =0 \qquad \forall x,y,z\in \frA
\end{equation}
where
\begin{equation}\label{eq:second-order leibniz identity}
    \Phi^3_D(x,y,z) \coloneqq \Phi^2_D(x,yz) - \Phi^2_D(x,y)z - (-1)^{(|D|+|x|)|y|}y\Phi^2_D(x,z).
\end{equation}
That is, the operator  \(\Phi^2_D(x,-)\) is a graded derivation. Equivalently, it is a shifted Poisson bracket or   Gerstenhaber bracket. 

Similarly, a \emph{third-order differential operator} \(D\colon\frA\to\frA\) satisfies 
\begin{equation}
    \Phi^4_D(x,y,z,w) =0 \qquad \forall x,y,z,w\in \frA
\end{equation}
where
\begin{equation}\label{eq:third-order leibniz identity}
    \Phi^4_D(x,y,z,w) \coloneqq \Phi^3_D(x,y,zw) - \Phi^3_D(x,y,z)w - (-1)^{(|D|+|x|+|y|)|z|}z\Phi^3_D(x,y,w).
\end{equation}
That is,  the operator  \(\Phi^3_D(x,y, -)\) is a graded derivation.

More generally, a differential operator \(D\colon\frA\to\frA\) is said to be of order at most $r-1$ if
\begin{equation}
\Phi_D^r(x_1,\ldots,x_r)=0
\qquad
\forall,x_1,\ldots,x_r\in\frA,
\end{equation}
where the higher \emph{Koszul brackets}  \(\Phi^r_D(x_1,\dotsc,x_r)\) are defined recursively by
\begin{multline}\label{eq:higher-order leibniz identity}
    \Phi^r_D(x_1,\dotsc,x_r) \coloneqq \Phi^{r-1}_D(x_1,\dotsc,x_{r-2},x_{r-1}x_r) - \Phi^{r-1}_D(x_1,\dotsc,x_{r-2},x_{r-1})x_r\\ - (-1)^{(|D|+|x_1|+\dotsb+|x_{r-2}|)|x_{r-1}|}x_{r-1}\Phi^{r-1}_D(x_1,\dotsc,x_{r-2},x_r)\,.
\end{multline}
Equivalently, for fixed $x_1,\ldots,x_{r-2}$, the operator
$\Phi_D^{r-1}(x_1,\ldots,x_{r-2},-)$ is a graded derivation. Note that an operator of order at most $r$ is automatically of order at most $s$ for every $s\geq r$.

 Written  in terms of $D$, the Koszul brackets are given by
\begin{equation}\label{expandedKbras}
\Phi_D^r(x_1,\dotsc,x_r)
= \sum_{\varnothing \neq I \subseteq \{1,\dotsc,r\}}
(-1)^{|I|}\,\varepsilon(I;x)\,
\Bigl(D\!\!\prod_{i\in I} x_i\Bigr)\prod_{j\notin I} x_j\,,
\end{equation}
where $\varepsilon(I;x)$ is the sign generated by moving the factors indexed by $I$ to the front without change their relative order within $I$ (i.e.~an unshuffle).
Explicitly, the first few are as follows:
\begin{equation}
\begin{aligned}
\Phi_D^1(x) &= Dx, \\
\Phi_D^2(x,y) &= D(xy) - (Dx)\,y - (-1)^{|x|}\,x\,(Dy) \\
\Phi_D^3(x,y,z) &= D(xyz) - D(xy)\,z - (-1)^{|y||z|}\,D(xz)\,y - (-1)^{|x|(|y|+|z|)}\,D(yz)\,x \\ 
&\qquad + (Dx)\,yz + (-1)^{|x|}\,x\,(Dy)\,z + (-1)^{|x|+|y|}\,xy\,(Dz),\\
\Phi_D^4(x_1,x_2,x_3,x_4)
&= D(x_1 x_2 x_3 x_4)- D(x_1 x_2 x_3)\,x_4+\cdots \\
&\qquad 
   - (-1)^{|x_3||x_4|}\,D(x_1 x_2 x_4)\,x_3+ \cdots\\
&\qquad  \cdots - (-1)^{|x_1|+|x_2|+|x_3|}\,x_1 x_2 x_3\,(Dx_4).
\end{aligned}
\end{equation}
It is clear that each monomial corresponds to a rooted tree with a single $D$-marked edge, so that $\Phi_D^r$ is the signed sum over all such singly-$D$-decorated trees.

\paragraph{The Koszul hierarchy.}  Let us now assume that $D$ has degree $|D|=-1$ and squares to zero (i.e.\ $D^2=0$). Setting $\Phi_D^1\coloneqq D$, the family of Koszul brackets
\begin{equation}
{\Phi_D^1,\Phi_D^2,\Phi_D^3,\dotsc}
\end{equation}
defines an $L_\infty$-algebra: the condition $D^2=0$ implies the full hierarchy of homotopy Jacobi identities among the $\Phi_D^i$. This $L_\infty$-algebra is known as the \emph{Koszul hierarchy} associated with $D$ \cite{koszul1985crochet,Akman:1995tm,Bering:1996kw,Markl:2013pca}.

If $D$ is of finite order at most $r$, then
\(
\Phi_D^{r+1}=0
\) implies
\(\Phi_D^n=0\) for \(n>r\),
and the Koszul hierarchy truncates to an $L_\infty$-algebra with brackets
$\Phi_D^1,\ldots,\Phi_D^r$. In particular, if $D$ is of order at most two, then $\Phi_D^n=0$ for all $n>2$, and the Koszul hierarchy reduces to a strict differential graded Lie algebra with differential $\Phi_D^1=D$ and bracket $\Phi_D^2$. For higher-order $D$, the failure of $\Phi_D^2$ to satisfy the strict Jacobi identity is controlled by the higher Koszul brackets beginning with $\Phi_D^3$.

\subsection{Higher-order   colour--kinematics duality}

Suppose we are in the situation of \cref{ckdual}, namely, that we have a cubic perturbative quantum field theory  whose  colour-stripped dg commutative algebra $(\frA, \sfm_2, \sfd)$ is, in fact, a strict BV${}^\Box$-algebra with a degree $-1$ operator $\sfb$ squaring to zero and satisfying \eqref{hodge} and \eqref{2ndorder}. The latter is just the condition that $\sfb$ be a second-order differential operator, and the kinematic Lie bracket is nothing but the binary Koszul bracket associated with $\sfb$:
\begin{equation}
[-,-]_{\text{kin}}=\Phi_\sfb^2(-,-). 
\end{equation}
 Full colour--kinematics duality corresponds to a Koszul hierarchy that truncates at order two. 
 
Phrased as such, a natural notion of higher-order colour--kinematics duality becomes self-evident. Suppose that a cubic perturbative quantum field theory  whose  colour-stripped dg commutative algebra $(\frA, \sfm_2, \sfd)$ admits a degree $-1$ operator $\sfb$ squaring to zero and satisfying \eqref{hodge}. It is then said to satisfy $n$\textsuperscript{th}-order colour--kinematics duality if $\sfb$ is order $n$ with respect to $\sfm_2$. That is, $n$\textsuperscript{th}-order colour--kinematics duality corresponds to a \emph{kinematic} Koszul hierarchy that truncates  with non-trivial brackets up to arity $n$:
\begin{equation}
\Phi^1_\sfb, \Phi^2_\sfb,\dotsc, \Phi^n_\sfb.
\end{equation}
The \emph{kinematic} Koszul hierarchy is a \emph{kinematic} $L_\infty$-algebra, so that the kinematic higher Koszul brackets satisfy the homotopy Jacobi relations. Explicitly, the homotopy Jacobi relations $J_n = 0$ for the first few $n$ read as follows:
\begin{align}
J_1\colon\quad & \Phi_\sfb^1\bigl(\Phi_\sfb^1(x_1)\bigr) = 0\,, \\[4pt]
J_2\colon\quad & \Phi_\sfb^1\bigl(\Phi_\sfb^2(x_1,x_2)\bigr)
  + \Phi_\sfb^2\bigl(\Phi_\sfb^1(x_1),x_2\bigr)
  + (-1)^{|x_1|}\,\Phi_\sfb^2\bigl(x_1,\Phi_\sfb^1(x_2)\bigr) = 0\,, \\[4pt]
\begin{split}
J_3\colon\quad & \Phi_\sfb^1\bigl(\Phi_\sfb^3(x_1,x_2,x_3)\bigr)
  + \Phi_\sfb^3\bigl(\Phi_\sfb^1(x_1),x_2,x_3\bigr)
  + (-1)^{|x_1|}\,\Phi_\sfb^3\bigl(x_1,\Phi_\sfb^1(x_2),x_3\bigr) \\
  &\ + (-1)^{|x_1|+|x_2|}\,\Phi_\sfb^3\bigl(x_1,x_2,\Phi_\sfb^1(x_3)\bigr)
  + \Phi_\sfb^2\bigl(\Phi_\sfb^2(x_1,x_2),x_3\bigr) \\
  &\ + (-1)^{(|x_1|+1)|x_2|}\,\Phi_\sfb^2\bigl(x_2,\Phi_\sfb^2(x_1,x_3)\bigr)
  + (-1)^{|x_1|}\,\Phi_\sfb^2\bigl(x_1,\Phi_\sfb^2(x_2,x_3)\bigr) = 0\,,
\end{split}
\end{align}
and in general
\begin{equation}
J_n\colon\quad
\sum_{i+j=n+1}\ \sum_{\sigma\in\mathrm{Sh}(i,n-i)}
\chi(\sigma;x)\,
\Phi_\sfb^{\,j}\bigl(\Phi_\sfb^{\,i}(x_{\sigma(1)},\dotsc,x_{\sigma(i)}),
x_{\sigma(i+1)},\dotsc,x_{\sigma(n)}\bigr) = 0\,,
\end{equation}
where the inner sum runs over $(i,n-i)$-unshuffles and $\chi(\sigma;x)$ is the corresponding Koszul sign. We stress that these are tautological identities: the higher Koszul bracket
$\Phi_\sfb^r$ is \emph{defined}, via the derivation recursion
\eqref{eq:higher-order leibniz identity}, precisely so that $J_r$ holds
identically.  The non-trivial content of the
Koszul-hierarchy construction is that $\sfb^2=0$ alone guarantees the whole
recursively-defined tower closes into an $L_\infty$-algebra. From this point of view, one immediately  sees that the failure of the strict kinematic Jacobi identity may always be resolved into an $L_\infty$-algebra for any theory, as described in \cite{Borsten:2023ned}.

If $\sfb$ is of finite exact order $r$, then the homotopy Jacobi relations hold on the nose (i.e.~not merely up to a higher chain homotopy) at order $r+1$. The Koszul hierarchy terminates at finite order, and  higher-order colour--kinematics duality becomes a non-trivial\footnote{As opposed to infinite order $\sfb$; since the  Koszul hierarchy is tautological, this  case is always realised for any $\sfb^2=0$ in any theory.} structural statement regarding the diagrammatics of the off-shell  colour-stripped theory.

Let us specialise  to   $\sfb$-gauge, i.e.\ assume that $\sfb(x_i)=0$. Every term in which $\Phi_\sfb^1$ acts on a bare
input vanishes, so that 
\begin{equation}
\Phi_\sfb^2(x,y) = \sfb(xy)\,,
\end{equation}
for all $x,y$ in $\sfb$-gauge. 

Similarly,  $J_1$ and $J_2$  trivialise, and the first two non-trivial
homotopy Jacobi relations reduce to the following:
\begin{subequations}
\begin{align}
\begin{split}
J_3\colon\quad &
 \Phi_\sfb^1\bigl(\Phi_\sfb^3(x_1,x_2,x_3)\bigr)+ \Phi_\sfb^2\bigl(\Phi_\sfb^2(x_1,x_2),x_3\bigr) \\
&\ + (-1)^{|x_2||x_3|}\,\Phi_\sfb^2\bigl(\Phi_\sfb^2(x_1,x_3),x_2\bigr)\\
&+ (-1)^{|x_1|(|x_2|+|x_3|)}\,\Phi_\sfb^2\bigl(\Phi_\sfb^2(x_2,x_3),x_1\bigr) = 0\,,
\end{split} \\
\begin{split}\label{J4}
J_4\colon\quad &
\Phi_\sfb^1\bigl(\Phi_\sfb^4(x_1,x_2,x_3,x_4)\bigr) \\
&\ + \Phi_\sfb^3\bigl(\Phi_\sfb^2(x_1,x_2),x_3,x_4\bigr)
+ (-1)^{|x_2||x_3|}\,\Phi_\sfb^3\bigl(\Phi_\sfb^2(x_1,x_3),x_2,x_4\bigr) \\
&\ + (-1)^{|x_2||x_4|+|x_3||x_4|}\,\Phi_\sfb^3\bigl(\Phi_\sfb^2(x_1,x_4),x_2,x_3\bigr)\\
&+ (-1)^{|x_1|(|x_2|+|x_3|)}\,\Phi_\sfb^3\bigl(\Phi_\sfb^2(x_2,x_3),x_1,x_4\bigr) \\
&\ + (-1)^{|x_1|(|x_2|+|x_4|)+|x_3||x_4|}\,\Phi_\sfb^3\bigl(\Phi_\sfb^2(x_2,x_4),x_1,x_3\bigr) \\
&\ + (-1)^{|x_1|(|x_3|+|x_4|)+|x_2|(|x_3|+|x_4|)}\,\Phi_\sfb^3\bigl(\Phi_\sfb^2(x_3,x_4),x_1,x_2\bigr) \\
&\ + \Phi_\sfb^2\bigl(\Phi_\sfb^3(x_1,x_2,x_3),x_4\bigr)
+ (-1)^{|x_3||x_4|}\,\Phi_\sfb^2\bigl(\Phi_\sfb^3(x_1,x_2,x_4),x_3\bigr) \\
&\ + (-1)^{|x_2|(|x_3|+|x_4|)}\,\Phi_\sfb^2\bigl(\Phi_\sfb^3(x_1,x_3,x_4),x_2\bigr) \\
&\ + (-1)^{|x_1|(|x_2|+|x_3|+|x_4|)}\,\Phi_\sfb^2\bigl(\Phi_\sfb^3(x_2,x_3,x_4),x_1\bigr) = 0\,.
\end{split}
\end{align}
\end{subequations}
In $\sfb$-gauge, the kinematic numerators of the Feynman diagrams are built exclusively from the vertex $\Phi^2_\sfb$ (with the root $\sfb$ stripped for off-shell amplitude numerators). Note that the image of $\Phi_\sfb^2$ is  $\sfb$-exact so that all internal edges of the kinematic numerators are in $\sfb$-gauge too (since we assume $\sfb^2=0$).

 Suppose that $\Phi^3_\sfb$ vanishes identically. In this case, if  a triple of diagrams $(\gamma_{J1},\gamma_{J2},\gamma_{J3})$ has colour numerators satisfying a colour Jacobi identity  $\sfc_{\gamma_{J1}}+\sfc_{\gamma_{J2}}+\sfc_{\gamma_{J3}}\ =\ 0$, then the corresponding kinematic numerators will satisfy an off-shell  graded kinematic Jacobi identity $ \sfn_{\gamma_{J1}}+\sfn_{\gamma_{J2}}+\sfn_{\gamma_{J3}}\ =\ 0$ due to $J_3$ on 
          \begin{equation}
            \gamma_{J1}= \begin{tikzpicture}[
                scale=1,
                every node/.style={scale=1},
                baseline={([yshift=-.5ex]current bounding box.center)}
                ]
                \matrix (m) [
                matrix of nodes,
                ampersand replacement=\&,
                column sep=0.13cm,
                row sep=0.13cm
                ]{
                    {} \& {}\& {} \& {} \& {}
                    \\
                    {} \& {} \& {}\& {} \& {}
                    \\
                    {} \& {} \& {} \& {} \& {}
                    \\
                    {} \& {} \& {}\& {} \& {}
                    \\
                    {} \& {} \& {} \& {} \& {}
                    \\
                };
                \draw [gluon] (m-1-1) -- (m-3-2.center);
                \draw [gluon] (m-5-1) -- (m-3-2.center);
                \draw [gluon] (m-3-2.center) -| node[near start,below] {} (m-3-4.center);
                \draw [gluon] (m-1-5) -- (m-3-4.center);
                \draw [gluon] (m-5-5) -- (m-3-4.center);
                \node [circle,draw,dashed,minimum size=1.9cm] (c) at (0,0){};
                \foreach \x in {(m-3-2), (m-3-4)}{
                    \fill \x circle[radius=2pt];
                }
            \end{tikzpicture},
            ~~~~~~
          \gamma_{J2}=  \begin{tikzpicture}[
                scale=1,
                every node/.style={scale=1},
                baseline={([yshift=-.5ex]current bounding box.center)}
                ]
                \matrix (m) [
                matrix of nodes,
                ampersand replacement=\&,
                column sep=0.13cm,
                row sep=0.13cm
                ]{
                    {} \& {} \& {} \& {} \& {}
                    \\
                    {} \& {} \& {} \& {} \& {}
                    \\
                    {} \& {} \& {} \& {} \& {}
                    \\
                    {} \& {} \& {} \& {} \& {}
                    \\
                    {} \& {} \& {} \& {} \& {}
                    \\
                };
                \draw [gluon] (m-1-1) -- (m-2-3.center);
                \draw [gluon] (m-5-1) -- (m-4-3.center);
                \draw [gluon] (m-2-3.center) -| node[near end,left] {} (m-4-3.center);
                \draw [gluon] (m-1-5) -- (m-2-3.center);
                \draw [gluon] (m-5-5) -- (m-4-3.center);
                \node [circle,draw,dashed,minimum size=1.9cm] (c) at (0,0){};
                \foreach \x in {(m-2-3), (m-4-3)}{
                    \fill \x circle[radius=2pt];
                }
            \end{tikzpicture},
            ~~~~~~
            \gamma_{J3}= \begin{tikzpicture}[
                scale=1,
                every node/.style={scale=1},
                baseline={([yshift=-.5ex]current bounding box.center)}
                ]
                \matrix (m) [
                matrix of nodes,
                ampersand replacement=\&,
                column sep=0.13cm,
                row sep=0.13cm
                ]{
                    {} \& {}\& {} \& {} \& {}
                    \\
                    {} \& {} \& {}\& {} \& {}
                    \\
                    {} \& {} \& {}\& {} \& {}
                    \\
                    {} \& {} \& {}\& {} \& {}
                    \\
                    {} \& {}\& {} \& {} \& {}
                    \\
                };
                \draw [gluon] (m-1-1) -- (m-3-4.center);
                \draw [gluon] (m-5-1) -- (m-3-2.center);
                \draw [gluon] (m-3-2.center) -| node[near start,below] {} (m-3-4.center);
                \draw [gluon] (m-1-5) -- (m-3-2.center);
                \draw [gluon] (m-5-5) -- (m-3-4.center);
                \node [circle,draw,dashed,minimum size=1.9cm] (c) at (0,0){};
                \foreach \x in {(m-3-2),(m-3-4)}{
                    \fill \x circle[radius=2pt];
                }
            \end{tikzpicture}.
        \end{equation}
If, on the other hand,  $\Phi^3_\sfb\not=0$,  then  $ \sfn_{\gamma_{J1}}+\sfn_{\gamma_{J2}}+\sfn_{\gamma_{J3}}\ =\ 0$ will fail. The failure will be given by $J_3$:
\begin{equation}\label{j3kin}
 \sfn_{\gamma_{J1}}+\sfn_{\gamma_{J2}}+\sfn_{\gamma_{J3}} ~\propto ~\Phi_\sfb^1\bigl(\Phi_\sfb^3(x_1,x_2,x_3)\bigr),
\end{equation}
where $x_1,x_2,x_3$ are inputs to subdiagrams (each possibly  roots of their own sub-trees) and $\Phi_\sfb^1\bigl(\Phi_\sfb^3(x_1,x_2,x_3)\bigr)$ is the output of the common root (which   itself may be a leaf on  another  sub-tree). If output is the root of the entire diagram and the $x_0$ is the final outstate of an (off-shell) amplitude then 
\begin{equation}
 \sfn_{\gamma_{J1}}+\sfn_{\gamma_{J2}}+\sfn_{\gamma_{J3}} ~\propto ~\langle \Phi_\sfb^3(x_1,x_2,x_3), x_0\rangle,
\end{equation}
so $\Phi_\sfb^3(x_1,x_2,x_3)$ itself is the failure of CK-duality. Note that this follows from the definition \eqref{eq:second-order leibniz identity} and  the observation that  $\langle \sfb(x_1x_2x_3), x_0\rangle=0$ for $x_0$ in $\sfb$-gauge.  

Recall that \eqref{j3kin}  is a tautological expression and so effectively content free. However,   if the Koszul hierarchy terminates at finite order, then we eventually hit a non-trivial relation amongst diagrams built from the colour-stripped Feynman vertices, but these diagrams are \emph{not} necessarily the kinematic numerators themselves.

More specifically, if $\sfb$ has order $n-1$, then $\Phi^n_\sfb = 0$ identically, and the defining term $\Phi^1_\sfb\bigl(\Phi^n_\sfb(-)\bigr)$ of $J_n$ drops out. The relation $J_n = 0$ thereby ceases to be tautological and becomes a genuine constraint that is bilinear in the lower brackets $\Phi^j_\sfb\circ\Phi^i_\sfb$ with $i+j=n+1$ and $2 \le i \le n-1$, where $\Phi^r_\sfb$ is the signed sum over $r$-leaf rooted trees (with vertices given by the colour-stripped Feynman vertex) carrying a single $\sfb$-marked edge (see \eqref{expandedKbras}). In $\sfb$-gauge, $\sfb$ annihilates the external legs, so any term whose mark lands on a leaf edge vanishes; the surviving terms are those in which the single $\sfb$-mark sits on an internal edge or the outgoing root edge. Thus, the identity is the signed sum over $n$-leaf binary rooted trees carrying two $\sfb$-insertions (that lie on the  internal lines or the root), in a nested configuration: one $\sfb$ sits on edge $e$, and the second lies strictly above $e$ on the rooted subtree that $e$ cuts out. 

 Finally, note that, since each Koszul bracket $\Phi^k_\sfb$ is the signed sum over rooted binary $\sfm_2$-trees (i.e.~colour-stripped Feynman vertices) carrying a single $\sfb$-decoration on one edge (i.e.~the numerator of the propagator), the finiteness condition $\Phi^n_\sfb=0$ implies that  the signed sum over all rooted binary trees on $n$ inputs, each with one $\sfb$ inserted on an internal or root edge (the leaf-edge insertions vanishing in $\sfb$-gauge), is required to cancel identically.

\section{Yang--Mills theory and higher-order colour--kinematics duality}\label{sec:main}
Having described the generic notion of higher-order CK-duality, we now demonstrate that the first standard first-order formalism of Yang--Mills theory with a \emph{single} auxiliary field admits a sixth-order differential operator $\sfb$. This should be contrasted with (super) Yang--Mills theory and M2-brane world-volume theories, which admit order two differential via  an infinite tower of auxiliary fields that are geometrised via twistor space or pure-spinor space  \cite{Ben-Shahar:2021doh,Borsten:2022vtg, Borsten:2023reb, Borsten:2023ned, Borsten:2023paw}. Also note that   the  standard first-order formalism of Yang--Mills theory in two dimensions admits an order two differential \cite{Ben-Shahar:2024dju}.

\subsection{First-order Batalin--Vilkovisky action of Yang--Mills theory}

Let us start off with a brief exposition of the first-order formulation of Yang--Mills theory. 
Take \((M,g)\) to be the (pseudo-)Riemannian spacetime manifold of dimension \(d\). 
We take the gauge group to be \(G\), and the corresponding gauge algebra to be \(\frg \coloneq \Lie(G)\). For simplicity, we take the gauge principal bundle to be trivial, \(P_G \cong M \times G\), and perform perturbation theory with respect to the trivial vacuum. We further take the gauge Lie algebra \(\frg\) to be equipped with a non-degenerate, \(G\)-invariant, symmetric form \(\kappa\colon \frg\otimes\frg\to\bbR\). Spacetime indices are denoted \(\mu,\nu,\cdots = 1,\cdots,d\) and Lie algebra indices are denoted \(a,b,\cdots=1,\cdots,\dim\frg\). For a generic basis \(\{T_a\}\) of \(\frg\) we take the structure constants to be \([T_a,T_b] = f_{ab}{^c}T_c\).

The second-order Yang--Mills action then reads
\begin{equation}
    S = \frac1{g_\ym^2}\int -\tfrac12 F \wedge\star F
    \coloneq \frac1{g_\ym^2}\int -\tfrac14\vol_g \, \kappa_{ab}F_{\mu\nu}^aF^{b\mu\nu}\,,
\end{equation}
where \(g_\ym^2\) a Yang--Mills coupling constant of mass dimension \(4-d\) and we defined the field strength \(F = \extd A + \frac12[A,A]\) in terms of the gauge connection \(A\).
This action contains quartic interactions and is described by a cyclic \(L_\infty\)-algebra with up to ternary brackets.
However, one can introduce a \((d-2)\)-form field \(B\) that transforms according to the adjoint representation of the gauge group such that the action with the auxiliary field included, namely
\begin{equation}
    S = \int F \wedge B +
    \tfrac12g_\ym^2B\wedge\star B\,,
\end{equation}
only contains cubic interactions; this is the first-order formulation of Yang--Mills theory \cite{Okubo:1979gt,McKeon:1994ds,Accardi:1997ps,Martellini:1997mu,Cattaneo:1997eh,Brandt:2015nxa,Frenkel:2017xvm,Brandt:2018avq,Brandt:2018wxe,Lavrov:2021pqh}, and it is described by a cyclic dg Lie algebra, with only the unary and binary brackets.

The Batalin--Vilkovisky action of Yang--Mills theory in the first-order formulation is given by 
\begin{equation}
\label{BV action}
\begin{aligned}
	S_\BV
	=
	\int &B\wedge\big(\extd A + \tfrac12 [A,A]\big) + \tfrac12g_\ym^2B\wedge\star B
	\\
	&- \big(\extd c + [A,c]\big)\wedge A^+ + [c,B]\wedge B^+ + \tfrac{1}{2}[c,c]\wedge c^+\,.
\end{aligned}
\end{equation}
Here, we introduced a scalar ghost field \(c\) of ghost number 1, and the corresponding antifields, namely a \((d-1)\)-form \(A^+\) and \(2\)-form \(B^+\) of ghost numbers \(-1\) and a \(d\)-form ghost antifield \(c^+\) of ghost number \(-2\), all in the adjoint representation of the gauge group. We refer the reader to \ref{ssec:BV-BRST conventions} for BV-BRST conventions.

\subsection{A superspace formulation}
\label{ssec:A superspace formulation of Yang--Mills theory}

The colour-stripped dg commutative algebra \(\frA\) of first-order Yang--Mills theory has a convenient superspace description. Consider the `thickened de Rham complex' \(\Omega_\theta^\bullet(M)\); a \(\bbZ\)-graded-commutative associative algebra
\begin{multline}
\label{eq:thickened de Rham complex}
	\Omega_\theta^\bullet(M) \coloneq \Omega^\bullet(M)[\theta]/(\theta^2) \cong \Omega^\bullet(M) \oplus \Omega^\bullet(M)\theta =
	\\
	\underbrace{
	\vsumb{\Omega^0(M)\theta}
	\rightarrow
	\cdots
	\rightarrow
	\overset c{\vsum{\Omega^0(M)}{\Omega^{d-3}(M)\theta}}
	}_{\text{excess}}
	\rightarrow
	\underset B{\overset A{\vsum{\Omega^1(M)}{\Omega^{d-2}(M)\theta}
	}}
	\rightarrow
	\underset{A^+}{\overset{B^+}{\vsum{\Omega^2(M)}{\Omega^{d-1}(M)\theta}
	}}
	\rightarrow
	\overbrace{\underset{c^+}{\vsum{\Omega^3(M)}{\Omega^d(M)\theta}
	}
	\rightarrow
	\cdots
	\rightarrow
	\vsumt{\Omega^d(M)}
	}^{\text{excess}}
\end{multline}
obtained by adjoining a degree-\((3-d)\) nilpotent generator \(\theta\) to the de Rham complex \(\Omega^\bullet(M)\). This complex contains the colour-stripped field content of Yang--Mills theory at the correct \(L_\infty\)-degree, alongside `excess' field content, as highlighted in \eqref{eq:thickened de Rham complex}. It should be emphasised that, while we use the same notation for the BV field content and their colour-stripped counterparts, the colour-stripped field content is \emph{not} valued in the gauge Lie algebra.

To rid ourselves of this excess field content, we first note that we can restrict ourselves to a homogeneous subalgebra
\begin{equation}
\label{eq:homogeneous subalgebra}
	\tilde\frA
	=
	\overset c{\vsumt{\Omega^0(M)}}
	\rightarrow 
	\underset B {\overset A{\vsum{\Omega^1(M)}{\Omega^{d-2}(M)\theta}
	}}
	\rightarrow
	\underset{A^+}{\overset {B^+}{\vsum{\Omega^2(M)}{\Omega^{d-1}(M)\theta}
	}}
	\rightarrow
	\overbrace{
	\underset{c^+}{\vsum{\Omega^3(M)}{\Omega^d(M)\theta}
	}
	\rightarrow
	\cdots
	\rightarrow
	\vsumt{\Omega^d(M)}
	}^{\text{homogeneous ideal $\frI$}}
	\,,
\end{equation}
removing one half of the excess field content. This subalgebra in turn admits a homogeneous ideal \(\frI = \Omega^{\bullet\geq 3}(M)\) consisting of the other half of the excess field content. Thus, by quotienting by this ideal, we obtain the colour-stripped Batalin--Vilkovisky complex of the first-order formulation of Yang--Mills theory,
\begin{equation}\label{eq:quotient-superspace}
	\frA \cong \tilde\frA/\frI
	\cong
	\overset c{\vsumt{\Omega^0(M)}}
	\rightarrow 
	\underset B {\overset A{\vsum{\Omega^1(M)}{\Omega^{d-2}(M)\theta}
	}}
	\rightarrow
	\underset{A^+}{\overset {B^+}{\vsum{\Omega^2(M)}{\Omega^{d-1}(M)\theta}
	}}
	\rightarrow
	\underset{c^+}{\vsumb{\Omega^d(M)\theta}}\,.
\end{equation}
This indeed agrees with the colour-stripping reviewed in \ref{ssec:Colour-stripping first order Yang-Mills theory} under the isomorphism \(c\mapsto c\), \(A\oplus B\mapsto A + B\theta\), \(A^+\oplus B^+\mapsto B^+ + A^+\theta\) and \(c^+ \mapsto c^+\theta\).

As for the differential on \(\frA\), in the superspace description it reads 
\begin{equation}
\label{Q superspace representation}
	\sfQ = \extd + g_\ym^2\pder{}{\theta}\star \sfP_{\Omega^{d-2}(M)\theta}\,,
\end{equation}
where \(\sfP_{\Omega^{d-2}(M)\theta}\) is the projection onto the subspace \(\Omega^{d-2}(M)\theta\) (cf.\ \ref{ssec:Projectors as differential operators}). Since \(\sfQ\) preserves both the homogeneous subalgebra \(\tilde\frA\) and homogeneous ideal \(\frI\), it descends to a well-defined operator on \(\frA\).
It is easy to see that \(\sfQ\) carries degree \(1\) and descends to a square-zero derivation on \(\frA\), which in fact agrees with the colour-stripped differential \eqref{eq:colour-stripped differential in components}.
If \(\mathfrak g\) is a simple Lie algebra, then \(\frL\coloneqq\frA\otimes\frg\) is a differential graded Lie algebra, whose homotopy Maurer--Cartan action reproduces the Batalin--Vilkovisky action \eqref{BV action} (cf.\ \ref{ssec:Colour-stripping first order Yang-Mills theory}). 

\subsection{The differential operator \(\sfb\)}
As explained in \cref{sec:review}, a weak form of colour--kinematics duality is implied by the existence of an operator \(\sfb\) that is a differential operator of low order. Consider the operator
\begin{equation}\label{eq:b-tilde definition}
	\tilde\sfb
	=
	\sfP_\frA\extd^\dagger\sfP_\frA - g_\ym^2\tpder{}\theta\star\sfP_{\Omega^{d-1}(M)\theta} + g_\ym^{-2}\theta\star^{-1}\extd^\dagger\extd\sfP_{\Omega^2(M)}
\end{equation}
defined on the thickened de Rham complex \(\Omega^\bullet_\theta(M)\), where \(\sfP_{(-)}\) are projectors onto their respective vector subspaces of \(\Omega_\theta^\bullet(M)\), elaborated on in \ref{ssec:Projectors as differential operators}. This operator descends to a well-defined operator on the subalgebra \(\tilde\frA\) (since \(\tilde\sfb\tilde\frA \subseteq \tilde\frA\)) and a well-defined operator on the subquotient algebra \(\frA\) (since \(\tilde\sfb\) preserves the homogeneous ideal \(\frI\), i.e.\ \(\tilde\sfb\frI \subseteq \frI\)).

This operator carries degree \(-1\) and is a differential operator of order \(4\) (cf.\ \ref{ssec:Order of sfb}). It satisfies \eqref{hodge} (cf.\ \cref{Appendix:Q-b-commute}), but does not square to zero: \(\tilde\sfb^2\ne0\). To repair this issue, we note that \eqref{hodge} is preserved by a \([\sfQ,-]\)-exact deformation of \(\tilde\sfb\). Accordingly, define
\begin{align}\label{eq:b definition}
    \sfb &= \tilde\sfb+[\sfQ,\sfw]\,,
    &
    \sfw &= K\intprod\big(\sfP_{\Omega^2(M)}\mp\sfP_{\Omega^d(M)\theta}\big)\,,
\end{align}
where \(\sfw\) is a degree-\((-2)\) map deforming the map \(\tilde\sfb\), and \(K\) is a 2-form normalised to square norm \(\pm 1\). This form is further taken to commute with d'Alembertian \(\Box\), in the sense that the following conditions are imposed,
\begin{align}\label{eq:K conditions}
	K\intprod K &= \pm 1\,,
	&
	[\Box,K\intprod]\sfP_{\Omega^2(M)} &= 0\,,
	&
	[\Box,K\wedge]\sfP_{\Omega^0(M)} &= 0\,,
\end{align}
where we have defined
\begin{align}
	K\wedge &= \tfrac12 K_{\mu\nu}\extd x^\mu\wedge\extd x^\nu\wedge\,,
	&
	K\intprod &= \tfrac12 K^{\mu\nu}\pder{}{\extd x^\nu}\pder{}{\extd x^\mu}\,.
\end{align}
The latter two conditions of \eqref{eq:K conditions} can be recast as
\begin{align}
	\Box K &= 0\,,
	&
	\nabla_{[\mu}K_{\nu]\rho} &= 0\,.
\end{align}
The second condition in particular implies that \(K\) is both closed and coclosed, since \((\extd K)_{\mu\nu\rho} = 3\nabla_{[[\mu}K_{\nu]\rho]}\) and \((\extd^\dagger K)_\nu = 2g^{\mu\rho}\nabla_{[\mu}K_{\nu]\rho}\). Such 2-forms always exist on flat space, where \(K\) can be taken to be constant.

A tedious but straightforward computation then shows that \(\sfb^2=0\) and that \(\sfb\) is a sixth-order differential operator; the details of the computation are given in \ref{ssec:Square of sfb vanishes} and \ref{ssec:Order of sfb}.

\section{Conclusions}

We have introduced a notion of \emph{higher-order} colour--kinematics duality, in
which the second-order differential operator $\sfb$ underlying ordinary colour--kinematics duality is
replaced by a differential operator of finite order $r > 2$. The associated Koszul
hierarchy endows the colour-stripped fields with a kinematic $L_\infty$-algebra whose
brackets $\Phi^1_\sfb,\dotsc,\Phi^r_\sfb$ are generated by $\sfb$ and the three-point
vertex. Generically this hierarchy is tautological, in the sense that each higher bracket
is defined precisely so as to satisfy the homotopy Jacobi relations. However, for finite order $n$ differentials $\Phi^n_\sfb=0$
imposes non-trivial constraints on the off-shell diagrams that may be formed by the colour-stripped Feynman vertices and propagator numerator. In $\sfb$-gauge these constraints take the form of
signed sums over tree diagrams with a single $\sfb$-insertion  that vanish identically and, being off-shell, they
imply that any  higher-point or loop-level diagrams containing them vanish equally. We then
demonstrated the existence of such structures  for pure Yang--Mills theory. Specifically, we demonstrated that a  superspace geometric formulation of the
first-order action of pure Yang--Mills theory in arbitrary spacetime dimension admits a
degree-$(-1)$ operator $\tilde\sfb$ of order four, which we repaired by a
$[\sfQ,-]$-exact deformation to a operator $\sfb$ of order six that squares to zero, thereby
exhibiting sixth-order colour--kinematics duality for pure Yang--Mills theory in any dimension using a single auxiliary field.

Several questions present themselves.
First, we have confined ourselves throughout to pure, non-supersymmetric Yang--Mills theory. It
is natural to ask whether the higher-order colour--kinematics duality established here
persists in the presence of matter or supersymmetry, and, if so, whether the order of
the corresponding operator $\sfb$ is raised, lowered, or left unchanged.  A second, closely related, structural question is whether the order of $\sfb$ can be reduced
 by the introduction of auxiliary fields. The examples of self-dual
Yang--Mills and the M2-brane theory show that an \emph{infinite} tower of auxiliary
fields can reduce the order all the way to two; the question is whether a \emph{finite}
number of auxiliary fields can achieve, or approach, the same reduction for the
order-six operator constructed here. We suspect that no finite enlargement reduces the
order to two (on the basis that the known approaches to manifesting tree-level colour--kinematics duality for Yang--Mills theory via an action principle involves all-order interactions), but we have not been able to rule it out. More
generally, one would like to understand which orders $r$ are realisable, and whether
there is a canonical minimal order attached to a given theory.

Finally,  our results raise the question of whether the
weaker fragment of colour--kinematics duality defined here suffices for some version of
the double copy. Ordinary double copy relies on the full second-order colour--kinematics duality; it is not
obvious whether an order-$n$ duality, which enforces only a subset of the kinematic
Jacobi identities exactly (with the remainder holding up to homotopy), is enough to
construct a consistent gravitational theory, or whether it instead yields a
correspondingly `homotopy-corrected' double copy in which the higher Koszul brackets
$\Phi^{\geq 3}_\sfb$ play an explicit role.  While intriguing, we leave this question to future work.

\appendix

\section{Conventions}\label{sec:Conventions}

\subsection{The Hodge star as a differential operator}
\label{ssec:The Hodge star as a differential operator}
We formulate the Hodge star operator \(\star\colon\Omega^\bullet(M) \to \Omega^\bullet(M)\) as a  higher order differential operators on the de Rham complex \(\Omega^\bullet(M)\).
Degree-wise, the Hodge star operator \(\star_p\colon\Omega^p(M)\to\Omega^{d-p}(M)\) acting on \(p\)-forms is given by
\begin{equation}\label{eq:hodge-star}
	\star_p\omega_p
	\coloneq
	\frac1{p!(d-p!)}\epsilon_{\mu_1\cdots\mu_d}\omega^{\mu_1\cdots\mu_p}\extd x^{\mu_{p+1}}\wedge\cdots\wedge\extd x^{\mu_d}\,,
\end{equation}
where the Levi-Civita tensor is normalised to \(\epsilon_{1\cdots d} = (\det g)\epsilon^{1\cdots d} = \sqrt{|\det g|}\). Its inverse $(\star_p)^{-1}: \Omega^{d-p}(M) \to \Omega^p(M)$ and  adjoint \((\star_p)^\dagger\colon\Omega^{d-p}(M) \to \Omega^p(M)\), the latter with respect to the inner product \(\braket{\omega}{\eta}_\dR \coloneq \int \omega\wedge\star\eta\), are given by \((\star^{-1})_p = (\det\eta)(\star^\dagger)_{d-p} = (\det\eta)(-1)^{p(d-p)}\star_{d-p}\).
As a differential operator of order \(p\) and degree \(d - 2p\) on \(\Omega^\bullet(M)\), the Hodge star operator \(\star_p\colon \Omega^\bullet(M)\to\Omega^\bullet(M)\) can be extended onto the entire de~Rham complex as
\begin{equation}
	\star_p 
	\coloneq 
	\frac1{p!(d-p!)}\epsilon^{\mu_1\cdots\mu_p}{_{\mu_{p+1}\cdots\mu_d}}
	\extd x^{\mu_{p+1}}\wedge\cdots\wedge\extd x^{\mu_d}\wedge\pder{}{\extd x^{\mu_p}}\cdots\pder{}{\extd x^{\mu_1}}\,.
\end{equation}
The kernel and image of these operators are given by \(\ker\star_p = \Omega^{\bullet\neq p}(M)\) and \(\im\star_p = \Omega^{d-p}(M)\) so that the Hodge star maps to the differential operator
\begin{equation}
	\star = \sum_{p=0}^d\star_p\,,
\end{equation}
with inverse 
\begin{equation}
\star^{-1} = (\det\eta)\star^\dagger = (\det\eta)(-1)^{\Upsilon(d-\Upsilon)}\star\,,
\end{equation}
where \(\Upsilon = \extd x^\mu \wedge\pder{}{\extd x^\mu}\) denotes the Euler vector field, which counts the form degree. The Hodge star is thus a differential operator of order \(d\) and of inhomogeneous degree.

\paragraph{Thickened de Rham complex.}
These operators are extended to the thickened de Rham complex \(\Omega^\bullet_\theta(M)\) introduced in \ref{ssec:A superspace formulation of Yang--Mills theory} through the graded Leibniz rule, together with \(\pder{}{x^\mu}\theta = \pder{}{\extd x^\mu}\theta \coloneq 0\). Note that the Euler vector field \(\Upsilon = \extd x^\mu\wedge\pder{}{\extd x^\mu}\) ignores the degree coming from \(\theta\). The degree-\((d-3)\) derivative \(\pder{}\theta\colon\Omega^\bullet_\theta(M) \to \Omega^\bullet_\theta(M)\) is defined by the graded Leibniz rule together with \(\ker\pder{}\theta = \Omega^\bullet(M)\) and \(\pder{}\theta\theta = 1\). In particular, it follows that \(\pder{^2}{\theta^2} = 0\). We still define \(\star^{-1} = (\det\eta)\star^\dagger = (\det\eta)(-1)^{\Upsilon(d-\Upsilon)}\star\) on \(\Omega_\theta^\bullet(M)\).

\subsection{Projectors as differential operators}
\label{ssec:Projectors as differential operators}

Using the properties of the Hodge star, we can also work out an expression of the de Rham projectors \(\sfP_p^{\mathrm{dR}}\colon\Omega^\bullet(M) \twoheadrightarrow \Omega^p(M)\). Namely, one can show that these operators are given by 
\begin{equation}
	\sfP_p^\dR 
	= 
	\star^{-1}\star_p
	=
	\sum_{k\geq p}\frac{(-1)^{k-p}}{p!(k-p)!}\extd x^{\mu_1}\wedge\cdots\wedge\extd x^{\mu_k}\wedge\pder{}{\extd x^{\mu_k}}\cdots\pder{}{\extd x^{\mu_1}}
\end{equation}
which follows from the fact that \(\ker\star_p = \Omega^{\bullet\neq p}(M)\) (cf.\ \ref{ssec:The Hodge star as a differential operator}), and then some differential form gymnastics. This differential operator is of de Rham degree \(0\) and order \(d\).
Let us now discuss projectors in the thickened de Rham complex \(\Omega^\bullet_\theta(M)\). We define the projectors
\begin{align}
	&
	\begin{aligned}
		\sfP_{\Omega^p(M)\phantom\theta} &\coloneq \pder{}{\theta}\theta\sfP^\dR_p 
		\colon
		\Omega^\bullet_\theta(M) \twoheadrightarrow \Omega^p(M)
		\\
		\sfP_{\Omega^p(M)\theta} &\coloneq \theta\pder{}{\theta}\sfP^\dR_p 
		\colon
		\Omega^\bullet_\theta(M) \twoheadrightarrow \Omega^p(M)\theta
	\end{aligned}
	\,,
	&
	\sfP_\frA &\coloneq \sum_{i=0}^2 \sfP_{\Omega^i(M)} + \sfP_{\Omega^{d-i}(M)\theta}\,,
\end{align}
where \(\sfP_p^\dR\) the natural extension of the de Rham projector to the thickened de Rham complex \(\Omega^\bullet_\theta(M)\). The projector \(\sfP_\frA\) projects onto the physical subspace \(\frA = \bigoplus_{i=0}^2\Omega^i(M) \oplus \Omega^{d-i}(M)\theta\) of the thickened de Rham complex \(\Omega^\bullet_\theta(M)\).

\subsection{de Rham codifferential and d'Alembertian}
\label{ssec:de Rham codifferential and d'Alembertian}

We define the de Rham codifferential \(\Omega^\bullet(M) \to \Omega^{\bullet-1}(M)\) to be the adjoint of the de Rham differential \(\extd\colon\Omega^\bullet(M)\to\Omega^{\bullet+1}(M)\) with respect to the inner product \(\braket--_\dR\) on the de Rham complex \(\Omega^\bullet(M)\). It is given by
\begin{equation}
	\extd^\dagger 
	=
	\star^{-1}\extd\star(-1)^\Upsilon
	=
	(-1)^{d-\Upsilon}\star\extd\star^{-1}
	=
	-\pder{}{\extd x^\mu}\nabla^\mu\,,
\end{equation}
where \(\nabla_\mu\) the covariant derivative with respect to the Levi-Civita connection. The de Rham codifferential is a differential operator of degree \(-1\) and order \(2\).
The d'Alembertian \(\Box\colon\Omega^\bullet(M) \to \Omega^\bullet(M)\) is defined to be the graded commutator between the de Rham differential and codifferential, and reads
\begin{equation}
	\Box 
	= 
	[\extd,\extd^\dagger]
	=
	-\nabla^\mu\nabla_\mu 
	+ R_\mu{^\nu}\extd x^\mu\wedge\pder{}{\extd x^\nu}
	+ \frac12 R_{\mu\nu}{^{\rho\sigma}}\extd x^\mu\wedge\extd x^\nu\wedge\pder{}{\extd x^\rho}\pder{}{\extd x^\sigma}\,,
\end{equation}
where \(R_{\mu\nu}{^{\rho\sigma}}\) the Riemann curvature tensor and \(R_\mu{^\nu} = R_{\mu\rho}{^{\nu\rho}}\) the Ricci tensor. It is a differential operator of degree \(0\) and order \(2\).
The de Rham differential, codifferential and d'Alembertian are extended to the thickened de Rham complex \(\Omega_\theta^\bullet(M)\) in the same way as outlined in \ref{ssec:The Hodge star as a differential operator}.

\subsection{Forms and fermions}
\label{ssec:Forms and fermions}

\paragraph{Functional-valued forms.}
Throughout this paper, we will consider differential forms which are valued in functionals on configuration space. Suppose we have some \(\bbZ\)-graded vector space \(\sfV\) (such as the Batalin--Vilkovisky configuration space), with functionals \(\cC^\infty(\sfV)\). Functional-valued polyforms are then given by
\(
	\Omega^\bullet(M,\cC^\infty(\sfV)) 
	\coloneq 
	\Omega^\bullet(M) \otimes \cC^\infty(\sfV)
\).
This space has two gradings: the de Rham grading \(|-|_\dR\) inherited from \(\Omega^\bullet(M)\), and the ghost grading \(|-|_\sfV\) inherited from \(\cC^\infty(\sfV)\). We follow Deligne's graded-commutation convention, \(\omega\wedge\eta = (-1)^{|\omega|_\dR|\eta|_\dR + |\omega|_\sfV|\eta|_\sfV}\eta\wedge\omega\), for \(\omega,\eta \in \Omega^\bullet(M,\cC^\infty(\sfV))\) elements of homogeneous degrees. 

\paragraph{Operations.}
Integration on functional-valued polyforms becomes a map \(\int\colon\Omega^\bullet(M,\cC^\infty(\sfV)) \to \cC^\infty(\sfV)\). Let \(v\) denote linear coordinate functions on \(\sfV\). We may extend the derivatives \(\pder{}{x},\pder{}{\extd x},\pder{}{v}\colon \Omega^\bullet(M,\cC^\infty(\sfV))\to\Omega^\bullet(M,\cC^\infty(\sfV))\) through the graded Leibniz rule and by setting \(\pder{}{x}v = \pder{}{\extd x}v = \pder{}{v}x = \pder{}{v}\extd x = 0\). Operators such as the Hodge star and de Rham projectors are then naturally extended to \(\Omega^\bullet(M,\cC^\infty(\sfV))\) by their differential operator expressions.

\subsection{BV-BRST conventions}
\label{ssec:BV-BRST conventions}

The infinitesimal gauge structure of first-order Yang--Mills theory is given by the Lie algebra of local gauge transformations \(\Omega^0(M,\frg)\). Accordingly, BRST configuration space is given by the degree-\((+1)\)-shifted action Lie algebroid
\begin{equation}
	\frF_\BRST 
	= \Omega^0(M,\frg)[1] \ltimes \frF_\ym
	= \Omega^0(M,\frg)[1]
	\rightarrow
	\vsum{\Omega^1(M,\frg)}{\Omega^{d-2}(M,\frg)}
	\,,
\end{equation}
on which the BRST differential \(Q_\BRST \in \Gamma(\tT\frF_\BRST)\) is given by
\begin{align}
	Q_\BRST A &= -\extd c - [A,c]
	\,,
	&
	Q_\BRST B &= [c,B]
	\,,
	&
	&
	Q_\BRST c = \tfrac12[c,c]\,.
\end{align}
Following \ref{ssec:Forms and fermions}, taking \(\sfV = \frF_\BRST\), forms commute with ghosts. We further denote \(|-|_\gh\) the \textit{ghost degree} on \(\cC^\infty(\frF_\BRST)\), so that \(|A|_\gh = |B|_\gh = 0\) and \(|c|_\gh = +1\). 
We obtain BV configuration space by taking the degree-\((-1)\)-shifted cotangent bundle of BRST configuration space,
\begin{equation}
\begin{split}
	\frF_\BV
	&\coloneq 
	\tT^\ast[-1]\frF_\BRST
	\\
	&=
	\overset c{\Omega^0(M,\frg)[1]}
	\rightarrow 
	\underset B{\overset A{\vsum{\Omega^1(M,\frg)}{\Omega^{d-2}(M,\frg)}
	}}
	\rightarrow 
	\underset{B^+}{\overset{A^+}{\vsum{\Omega^{d-1}(M,\frg)[-1]}{\Omega^2(M,\frg)[-1]}
	}}
	\rightarrow 
	\overset{c^+}{\Omega^d(M,\frg)[-2]}
\end{split}
\end{equation}
where we made use of the fact that \(\tT^\ast_0\Omega^p(M,\frg) \cong \Omega^{d-p}(M,\frg)\) using the invariant bilinear form \(\kappa\). We note that the collective fields and ghosts \((A,B,c)\) and their antifield counterparts \((A^+,B^+,c^+)\) have their ghost numbers related in that they pairwise add up to \(-1\).
We work in the convention where the BV antibracket \((-,-)_\BV\) on \(\cC^\infty(\frF_\BV)\) is characterised by
\begin{equation}
\begin{split}
	\big(A^a_{\mu_1}(x),A^{b+}_{\mu_2\cdots\mu_d}(y)\big)_\BV
	=
	\big(B^a_{\mu_1\mu_2}(x),B^{b+}_{\mu_3\cdots\mu_d}(y)\big)_\BV
	=
	-\big(c^a(x),c^{b+}_{\mu_1\cdots\mu_d}(y)\big)_\BV
	\\
	=
	\delta(x,y)\kappa^{ab}\varepsilon_{\mu_1\cdots\mu_d}
\end{split}
\end{equation}
antisymmetric in field-antifield pairs (where \(\varepsilon_{\mu_1\cdots\mu_d}\) is the Levi-Civita density of weight \(+1\), as opposed to the Levi-Civita tensor \(\epsilon_{\mu_1\cdots\mu_d}\) in \eqref{eq:hodge-star})
together with the symmetry and derivative properties
\begin{equation}
\begin{gathered}
	(F,G)_\BV = -(-1)^{(|F|_\gh+1)(|G|_\gh+1)}(G,F)\,,
	\\
	(F,GH)_\BV = (F,G)_\BV H + (-1)^{(|F|_\gh + 1)|G|_\gh}G(F,H)_\BV\,,
	\\
	(FG,H)_\BV = F(G,H)_\BV + (-1)^{(|H|_\gh + 1)|G|_\gh}(F,H)_\BV G\,.
\end{gathered}
\end{equation}
The BV action \(S_\BV\) is then characterised by the master equation \((S_\BV,S_\BV)_\BV = 0\), and reads
\begin{equation}
\label{eq:first order Yang-Mills BV action}
\begin{split}
	S_\BV
	&=
	S + \int Q_\BRST A\wedge A^+ + Q_\BRST B\wedge B^+ + Q_\BRST c\wedge c^+
	\\
	&= \int
	\begin{aligned}[t]
		&B\wedge\big(\extd A + \tfrac12[A,A]\big) + \tfrac12 g_\ym^2B\wedge\star B
		\bigg.\\
		&+ \big(\extd c + [A,c]\big)\wedge A^+
		- [c,B]\wedge B^+
		+ \tfrac12 [c,c]\wedge c^+\,,
		\bigg.
	\end{aligned}
\end{split}
\end{equation}
provided we impose boundary conditions \(S_\BV|_{\frF_\BRST} = S\) and \((S_\BV,-)_\BV|_{\frF_\BRST} = Q_\BRST\).

\subsection{Colour-stripping first-order Yang--Mills theory}\label{ssec:Colour-stripping first order Yang-Mills theory}

To derive the underlying \(L_\infty\)-algebra structure of the theory, we construct the homotopy Maurer-Cartan action \(S_\hMC\) of the theory. The underlying \(L_\infty\)-algebra is given by the degree-\((-1)\)-shifted BV configuration space \(\frL \coloneq \frF_\BV[-1] \Leftrightarrow \frF_\BV = \frL[1]\). We denote the corresponding basis \((\sfe,\sff,\sfg)\) and \((\sfe^+,\sff^+,\sfg^+)\) dual to respectively the fields \((A,B,c)\) and antifields \((A^+,B^+,c^+)\). These carry \(L_\infty\)-degrees \((|\sfe|_\frL,|\sff|_\frL,|\sfg|_\frL) = (1,1,0)\) and \((|\sfe^+|_\frL,|\sff^+|_\frL,|\sfg^+|_\frL) = (2,2,3)\). The degree-\((-1)\)-symplectic structure on \(\frF_\BV\) induces a degree-\((-3)\)-cyclic structure \(\braket--\colon \frL\otimes\frL\to\bbR\) on the underlying \(L_\infty\)-algebra \(\frL\), which reads
\begin{equation}
\begin{split}
	\braket{\sfe_a^{\mu_1}(x)}{\sfe^{+\mu_2\cdots\mu_d}_b(y)}
	=
	\braket{\sff_a^{\mu_1\cdots\mu_{d-2}}(x)}{\sff^{+\mu_{d-1}\mu_d}_b(y)}
	=
	-\braket{\sfg_a(x)}{\sfg^{+\mu_1\cdots\mu_d}_b(y)}
	\\
	=
	\kappa_{ab}\varepsilon^{\mu_1\cdots\mu_d}\delta(x,y)
\end{split}
\end{equation}
antisymmetric in basis-antibasis pairs (where \(\varepsilon^{\mu_1\cdots\mu_d}\) is the Levi-Civita density of weight \(-1\), as opposed to the Levi-Civita tensor \(\epsilon_{\mu_1\cdots\mu_d}\) in \eqref{eq:hodge-star}).
We further perform an extension of the cyclic structure to the \(\cC^\infty(\frF_\BV)\)-module \(\cC^\infty(\frF_\BV,\frL) \coloneq \cC^\infty(\frF_\BV)\otimes\frL\) according to Bernstein's conventions \(\braket{F\otimes\ell}{G\otimes\ell'} \coloneq (-1)^{|G|_\gh(|\ell|_\frL + 1)}FG\braket\ell{\ell'}\). 

\paragraph{Homotopy Maurer-Cartan action.} On this module we introduce the \textit{contracted coordinate function}
\begin{equation}
\begin{split}
	\sfa
	&\coloneq \int\extd^dx ~ \Big[
	\begin{aligned}[t]
		&A_\mu^a(x)\otimes\sfe^\mu_a(x) + A^{+a}_{\mu_1\cdots\mu_{d-1}}(x)\otimes\sfe^{+\mu_1\cdots\mu_{d-1}}_a(x)
		\Big.\\
		&+ B^a_{\mu_1\cdots\mu_{d-2}}(x)\otimes\sff_a^{\mu_1\cdots\mu_{d-2}}(x)
		+
		B^{+a}_{\mu\nu}(x)\otimes\sff^{+\mu\nu}_a(x)
		\Big.\\
		&+ c^a(x)\otimes\sfg_a(x) 
		+ 
		c^{+a}_{\mu_1\cdots\mu_d}(x)\otimes\sfg^{+\mu_1\cdots\mu_d}_a(x)
		\Big] 
	\end{aligned}
	\\
	&\eqcolon A + B + c + A^+ + B^+ + c^+\,.
\end{split}
\end{equation}
Written in terms of the cyclic structure and contracted coordinate function, the action takes on the form of the \emph{homotopy Maurer-Cartan action}
\begin{equation}
\label{eq:hMC action general}
	S_\hMC[\sfa]
	=
	\sum_{k=1}^\infty\tfrac1{(k+1)!}\braket\sfa{\hat\mu_k(\sfa,\cdots,\sfa)}\,,
\end{equation}
where the degree-\((2-k)\) maps \(\hat\mu_k\colon \bigwedge^i\cC^\infty(\frF_\BV,\frL) \to \cC^\infty(\frF_\BV,\frL)\) are the \(i\)-ary brackets defining the underlying \(\cC^\infty(\frF_\BV)\)-module \(L_\infty\)-algebra \((\cC^\infty(\frF_\BV,\frL),\{\hat\mu_i\}_{i=1}^\infty)\). From these, we can in turn read off the underlying \(L_\infty\)-algebra \((\frL,(\mu_i)_{i=1}^\infty)\), using Bernstein's conventions, as \(\hat\mu_k(F\otimes\ell,G\otimes\ell',\cdots) = (-1)^{k|F|_\gh + (k+|\ell|_\frL)|G|_\gh + \cdots}(FG\cdots)\otimes\mu_k(\ell,\ell',\cdots)\).
In our case, the only non-vanishing brackets will be \((\mu_1,\mu_2)\), forming a dg Lie algebra. Written as the homotopy Maurer-Cartan action \eqref{eq:hMC action general}, the first-order BV action \eqref{eq:first order Yang-Mills BV action} becomes
\begin{equation}
\begin{aligned}
	S_\hMC
	={}
	&\tfrac12 \braket B{\extd A + g_\ym^2 {\star}B} 
	+ \tfrac12\braket{A}{\extd B}
	- \tfrac12\braket c{\extd A^+}
	- \tfrac12\braket{A^+}{\extd c}
	\\
	&+ \tfrac1{3!}\braket B{[A,A]} + \tfrac13\braket A{[A,B]} + \tfrac13\braket A{[c,A^+]} - \tfrac13\braket c{[A,A^+]}
	\\
	&+ \tfrac13\braket{A^+}{[c,A]} + \tfrac13\braket B{[c,B^+]} - \tfrac13\braket c{[B,B^+]} + \tfrac13\braket{B^+}{[c,B]}
	\\
	&+ \tfrac1{3!}\braket{c^+}{[c,c]} + \tfrac13\braket{c}{[c,c^+]}\,.
\end{aligned}
\end{equation}
where we write \(\extd A = \int\extd^dx\,\partial_\mu A^a_\nu(x)\otimes\sff^{+\mu\nu}_a(x)\), and so forth.
Comparing this to \eqref{eq:hMC action general}, we can read off the unary bracket \(\mu_1\colon \frL_k \to \frL_{k+1}\) and the binary bracket \(\mu_2\colon \frL_k\wedge\frL_\ell\to\frL_{k+\ell}\), forming a dg Lie algebra. The unary bracket has non-vanishing components
\begin{align}
	\mu_1(c) &= \extd c \oplus 0\,,
	&
	\mu_1(A\oplus B) &= \extd B \oplus (\extd A + g_\ym^2{\star}B)\,,
	&
	\mu_1(A^+\oplus B^+) &= \extd A^+\,.
\end{align}
As for the binary bracket, we find
\begin{equation}
\begin{gathered}
	\begin{aligned}
	\mu_2(c_1,c_2) &= [c_1,c_2]\,,
	&
	\mu_2(c,A\phantom{^+}\oplus B)\phantom{^+} &= [c,A]\phantom{^+} \oplus [c,B]\,,
	\\
	\mu_2(c,c^+) &= [c,c^+]\,,
	&
	\mu_2(c,A^+\oplus B^+) &= [c,A^+] \oplus [c,B^+]\,,
	\end{aligned}
	\\
	\begin{aligned}
	\mu_2(A_1\oplus B_1,A_2\oplus B_2) 
	&= \big([A_1, B_2] - [A_2, B_1]\big) \oplus [A_1, A_2]\,,
	\\
	\mu_2(A\oplus B,A^+\oplus B^+)
	&= [A, A^+] + [B, B^+]\,,
	\end{aligned}
\end{gathered}
\end{equation}
with other components vanishing for degree reasons.
\paragraph{Colour-stripping the dg Lie algebra.} To colour-strip the theory, we factor out the colour Lie algebra \(\frg\) from the dg Lie algebra \((\frL,\mu_1,\mu_2)\). This yields the dg commutative associative algebra \((\frA,\sfQ,\sfm)\),
\begin{equation}
	\frA \coloneq \Omega^0(M) \to \vsum{\Omega^1(M)}{\Omega^{d-2}(M)} \to \vsum{\Omega^{d-1}(M)}{\Omega^2(M)} \to \Omega^d(M)
\end{equation}
with the differential \(\sfQ\colon \frA\to\frA\) given by
\begin{align}
\label{eq:colour-stripped differential in components}
	\sfQ(c) &= \extd c \oplus 0\,,
	&
	\sfQ(A\oplus B) &= \extd B \oplus (\extd A + g_\ym^2{\star}B)\,,
	&
	\sfQ(A^+\oplus B^+) &= \extd A^+\,.
\end{align}
The nonzero components of the degree-\(0\) graded commutative associative product \(\sfm\colon \frA\otimes\frA\to\frA\) are
\begin{equation}
\begin{gathered}
	\begin{aligned}
	\sfm(c_1,c_2) &= c_1c_2\,,
	&
	\sfm(c,A\phantom{^+}\oplus B)\phantom{^+} &= cA\phantom{^+} \oplus cB\,,
	\\
	\sfm(c,c^+) &= cc^+\,,
	&
	\sfm(c,A^+\oplus B^+) &= cA^+ \oplus cB^+\,,
	\end{aligned}
	\\
	\begin{aligned}
	\sfm(A_1\oplus B_1,A_2\oplus B_2) 
	&= \big(A_1\wedge B_2 - A_2\wedge B_1\big) \oplus \big(A_1\wedge A_2\big)\,,
	\\
	\sfm(A\oplus B,A^+\oplus B^+)
	&= A\wedge A^+ + B\wedge B^+\,.
	\end{aligned}
\end{gathered}
\end{equation}
It is important to note that, while we still use the same notation for the field content after colour-stripping, these fields have no colour indices or ghost degree.

\section{Computations}\label{sec:Computations}

\subsection{Graded commutator of \(\sfb\) and \(\sfQ\)}\label{Appendix:Q-b-commute}
Let us show that \([\sfb,\sfQ]=\Box\) as claimed.
In the expression \eqref{eq:b definition}, the term \([\sfQ,\sfw]\) automatically commutes with \(\sfQ\). Thus it suffices to show that \([\tilde\sfb,\sfQ]=\Box\). The operators \(\sfQ\) and \(\tilde\sfb\) have non-vanishing components 
\begin{equation}
	\Omega^0(M) 
	\quad
	\overset{\sfQ_0}{\underset{\tilde\sfb_1}\rightleftarrows}
	\quad
	\vsum{\Omega^1(M)}{\Omega^{d-2}(M)}
	\quad
	\overset{\sfQ_1}{\underset{\tilde\sfb_2}\rightleftarrows}
	\quad
	\vsum{\Omega^{d-1}(M)}{\Omega^2(M)}
	\quad
	\overset{\sfQ_2}{\underset{\tilde\sfb_3}\rightleftarrows}
	\quad
	\Omega^d(M)\,,
\end{equation}
which read
\begin{align}
	\sfQ_0 &=
	\begin{pmatrix}
		\extd \\ 0
	\end{pmatrix}
	&
	\sfQ_1 &=
	\begin{pmatrix}
		0 & \extd
		\\
		\extd & g_\ym^2\star
	\end{pmatrix}
	&
	\sfQ_2 &=
	\begin{pmatrix}
		\extd & 0
	\end{pmatrix}\,,
	\\
	\tilde\sfb_1 &= 
	\begin{pmatrix}
		\extd^\dg & 0
	\end{pmatrix}
	&
	\tilde\sfb_2 &=
	\begin{pmatrix}
		-g_\ym^2\star^\dagger & \extd^\dg
		\\
		\extd^\dg & 
		g_\ym^{-2}
		\star^{-1}\extd^\dg\extd
	\end{pmatrix}
	&
	\tilde\sfb_3 &=
	\begin{pmatrix}
		\extd^\dg \\ 0
	\end{pmatrix}\,.
\end{align}
A direct computation then shows that \(\sfQ\tilde\sfb+\tilde\sfb\sfQ=\Box\) as claimed.

\subsection{Square of \(\sfb\) vanishes}
\label{ssec:Square of sfb vanishes}
\newcommand{\bsfb}{\bar{\sfb}}
\newcommand{\bdel}{\bar{\del}}

The operator \(\sfw\colon\frA_\bullet \to \frA_{\bullet-2}\) has non-vanishing components
\begin{align}
	\sfw_2 &= 
	\begin{pmatrix}
		0 & K\intprod
	\end{pmatrix}
	\colon
	\vsum{\Omega^{d-1}(M)}{\Omega^2(M)} \to \Omega^0(M)\,,
	&
	\sfw_3 &=
	\begin{pmatrix}
		0
		\\
		\mp K\intprod
	\end{pmatrix}
	\colon
	\Omega^d(M) \to \vsum{\Omega^1(M)}{\Omega^{d-2}(M)}\,.
\end{align}
Accordingly, the non-vanishing components of the operator \(\sfb \coloneq \tilde\sfb + [\sfQ,\sfw]\) read
\begin{align}
	\sfb_1 &=
	\begin{pmatrix}
		\extd^\dagger - K\intprod\extd
		\\
		-g_\ym^2K\intprod\star
	\end{pmatrix}^\top\,,
	&
	\sfb_2 &= 
	\begin{pmatrix}
		-g_\ym^2\star^\dagger & \extd^\dagger + \extd(K\intprod)
		\\
		\extd^\dagger \pm (K\intprod)\extd & g_\ym^{-2}\star^{-1}\extd^\dagger\extd
	\end{pmatrix}\,,
	&
	\sfb_3 &=
	\begin{pmatrix}
		\extd^\dagger \mp \extd(K\intprod)
		\\
		\mp g_\ym^2 \star (K \intprod)
	\end{pmatrix}\,,
\end{align}
where \((-)^\top\) denotes the transpose. Accordingly, the only components of \(\sfb^2\) which don't vanish for degree reasons are given by
\begin{gather}
	\sfb_1\sfb_2 
	=
	\begin{pmatrix}
		g_\ym^2\big(1 \mp K\intprod K\big){\star}\extd
		\\
		[\Box,K\intprod]
	\end{pmatrix}^\top 
	\colon
	\vsum{\Omega^{d-1}(M)}{\Omega^2(M)}
	\to
	\Omega^0(M)
	\\
	\sfb_2\sfb_3 
	= 
	\begin{pmatrix}
		g_\ym^2\extd\big(1 \mp K\intprod K\big)\star
		\\
		\mp [\Box,K\intprod]
	\end{pmatrix}\phantom{^\top}
	\colon
	\Omega^d(M)
	\to
	\vsum{\Omega^1(M)}{\Omega^{d-2}(M)}
\end{gather}
It thus follows that \(\sfb^2 = 0\) if \eqref{eq:K conditions} are satisfied.

\subsection{Order of \(\sfb\)}
\label{ssec:Order of sfb}

Let us show that as differential operators on \(\frA\), \(\tilde\sfb\) is of order six on \(\frA\) and \([\sfQ,\sfw]\) is of order at most six, and consequently, that \(\sfb=\tilde\sfb+[\sfQ,\sfw]\) of order six. Throughout this section, we will make use of facts and notation introduced in \ref{ssec:The Hodge star as a differential operator}, \ref{ssec:Projectors as differential operators}, and \ref{ssec:de Rham codifferential and d'Alembertian}, as well as everything related to the Koszul hierarchy reviewed in \ref{ssec:Higher-order differential operators and the Koszul hierarchy}.

\paragraph{Order of \(\tilde\sfb\).} 
Because each term in \eqref{eq:b-tilde definition} appears with a different power of the Yang--Mills coupling constant \(g_\ym^2\), the order of \(\tilde\sfb\) will be the maximum of the orders of these terms:
\begin{equation}
\label{eq:order tilde sfb as maximum}
	\Order_\frA(\tilde\sfb)
	=
	\max\Big\{
	\Order_\frA\big(\sfP_\frA\extd^\dagger\sfP_\frA\big),
	\Order_\frA\big(\tpder{}{\theta}\star_{d-1}\big),
	\Order_\frA\big(\theta\extd\star_3\extd\big)
	\Big\}\,.
\end{equation}
Recall furthermore that the Koszul brackets \eqref{eq:higher-order leibniz identity} carry degree
\begin{equation}
\label{eq:brace degrees}
	|\Phi_D^r(x_1,\cdots,x_r)|
	=
	|D| + |x_1| + \cdots + |x_r|\,.
\end{equation}
Considering the fact that \(|\tilde\sfb| = -1\) and \(\frA\) is concentrated in degrees \(0,1,2,3\), it follows that the Koszul brackets vanish for degree reasons except for when the arguments \(x_1,\cdots,x_r\) are constrained in degree by
\begin{equation}
	1 \leq |x_1| + \cdots + |x_r| \leq 4\,.
\end{equation}
This leaves one with ten cases of \((|x_1|,\cdots,|x_r|)\), which read
\begin{equation*}
\begin{aligned}
	(\vec0,1)
	\,,
	(\vec0,2)
	\,,
	(\vec0,3)
	\,,
	(\vec0,1,1)
	\,,
	(\vec0,1,2)
	\,,
	(\vec0,1,3)
	\,,
	(\vec0,2,2)
	\,,
	(\vec0,1,1,1)
	\,,
	(\vec0,1,1,2)
	\,,
	(\vec0,1,1,1,1)
	\,,
\end{aligned}
\end{equation*}
where we denote \(\vec0 = (0,\cdots,0)\) a `tail of ghosts', which for large enough length kill the Koszul brackets and give a finite order. 
This is short enough to check manually.

Starting with the order of \(\sfP_\frA\extd^\dagger\sfP_\frA\), we note that for a lot of computations, the projectors \(\sfP_\frA\) won't play a role, and one will essentially be computing Koszul brackets \(\Phi^{k+1}\) for the de~Rham codifferential \(\extd^\dagger\) on the thickened de Rham complex \(\Omega^\bullet_\theta(M)\), which vanish at order \(k = 2\). Let us list the non-trivial cases in which the projectors do play a role:
\((\vec c,B\theta)\), \((\vec c,A,B\theta)\), \((\vec c,A,B^+)\), \((\vec c,A_1,A_2,A_3)\), \((\vec c,A_1,A_2,B\theta)\), \((\vec c,A,B^+,B\theta)\) and \((\vec c,A_1,A_2,A_3,B\theta)\), where we denote a tail of ghosts of length \(r\) by \(\vec c = (c_1,\cdots,c_r)\). Among these, the leading contribution to the order is given by \((\vec c,A_1,A_2,A_3,B\theta)\). Let $\sfr=\sfP_\frA\extd^\dagger\sfP_\frA\colon\frA\to \frA$. Repeated application of the higher-order Leibniz rule \eqref{eq:higher-order leibniz identity} gives
\begin{equation}
\begin{split}
	&\Phi^{r+4}_\sfr\Big(\pi(\vec c),\pi(A_1),\pi(A_2),\pi(A_3),\pi(B\theta)\Big)
	\\
	&= \frac1{3!}\sum_{\sigma\in S_3}\pi\Big(
	\begin{aligned}[t]
		&3\Phi^{r+1}_{\extd^\dagger} \big(\vec c,A_{\sigma(1)}\wedge A_{\sigma(2)}\big)\wedge A_{\sigma(3)}\wedge B\theta
		\\
		&- 3\Phi_{\extd^\dagger}^{r+1}\big(\vec c,A_{\sigma(1)}\big)\wedge A_{\sigma(2)}\wedge A_{\sigma(3)}\wedge B\theta
	\Big)\,.
	\end{aligned}
\end{split}
\end{equation}
Here, \(\pi\colon\tilde\frA\twoheadrightarrow\frA\) denotes the projection from the homogeneous subalgebra \(\tilde\frA \subset \Omega^\bullet_\theta(M)\) defined in \eqref{eq:homogeneous subalgebra} onto the superspace \(\frA \cong \tilde\frA/\frI\). We note that this expression only vanishes when \(r\geq 2\), because the de Rham codifferential is an order-two differential operator on the de Rham complex \(\Omega^\bullet(M)\). From this we conclude that \(\sfP_\frA\extd^\dagger\sfP_\frA\) is an order five differential operator on \(\frA\),
\begin{equation}
\label{eq:order first term}
	\Order_\frA\big(\sfP_\frA\extd^\dagger\sfP_\frA\big)
	=
	5\,.
\end{equation}

Next up is the order of the operator \(\sfs=\pder{}\theta\star_{d-1} = \pder{}\theta\star\sfP_{\Omega^{d-1}(M)\theta}\colon \frA\to\frA\). For the computation to be non-trivial, a product of the components of the argument \(\vec x\) of the Koszul bracket \(\Phi^k_{\pder{}{\theta}\star_{d-1}}(\vec x)\) has to lie in \(\Omega^{d-1}(M)\theta\). This leaves us with the following non-trivial cases for \(\vec x\): \((\vec c,A^+\theta)\), \((\vec c,A,B\theta)\), \((\vec c,A,A^+\theta)\), \((\vec c,B\theta,A^+\theta)\), \((\vec c,A_1^+\theta,A_2^+\theta)\), \((\vec c,A_1,A_2,B\theta)\), \((\vec c,A,B_1\theta,B_2\theta)\), \((\vec c,A,B\theta,A^+\theta)\), \((\vec c,A_1,A_2,B_1\theta,B_2\theta)\).
The leading contribution to the order will come from \((\vec c,A_1,A_2,B_1\theta,B_2\theta)\). Applying the higher-order Leibniz rule \eqref{eq:higher-order leibniz identity} we find that
\begin{equation}
\begin{split}
	&\Phi^{r+4}_\sfs\Big(\pi(\vec c),\pi(A_1),\pi(B_1\theta),\pi(A_2),\pi(B_2\theta)\Big)
	\\
	&= 
	\sum_{\sigma,\tau \in S_2}\varepsilon(\sigma)\varepsilon(\tau)\pi\Big(
	\Phi^{r+1}_{(\star^\dagger)_{d-1}}\big(\vec c,A_{\sigma(1)}\wedge B_{\tau(1)}\big)\wedge A_{\sigma(2)}\wedge B_{\tau(2)}\theta
	\Big)
\end{split}
\end{equation}
which vanishes when \(r \geq 1\) since
\begin{equation}
	\Phi^{r+1}_{(\star^\dagger)_{d-1}}\big(c,\omega_{d-1}\big)
	=
	\star^\dagger(c\omega_{d-1}) - c\star^\dagger\omega_{d-1}
	=
	0\,.
\end{equation}
From this, we conclude that the order of \(\pder{}{\theta}\star_{d-1}\) on \(\frA\) is four,
\begin{equation}
\label{eq:order second term}
	\Order_\frA\big(\tpder{}\theta\star_{d-1}\big) = 4\,.
\end{equation}

Finally, we look at the order of \(\theta\extd\star_3\extd\). We start off by noting that one can rewrite
\begin{equation}
\begin{split}
	&\extd \star_p\extd
	=
	\extd[\star_p,\extd]
	\\
	&=
	\tfrac{p}{(p-1)!(d-p+1!)}\epsilon_{\mu_1\cdots\mu_{p-1}}{^{\mu_p\cdots\mu_d}}
	\extd x^{\mu_p}\wedge\cdots\wedge\extd x^{\mu_d}\wedge
	\pder{}{\extd x^{[\mu_{p-1}}}\cdots\pder{}{\extd x^{\mu_1}}\nabla^\nu\nabla_{\nu]}\,.
\end{split}
\end{equation}
From this it follows that 
\begin{equation}
	\Order_{\Omega^\bullet(M)}(\extd\star_p\extd)
	=
	\Order_{\tilde\frA}(\theta\extd\star_p\extd)
	=
	p+1\,.
\end{equation}
However, we would like to compute the order of \(\theta\extd\star_3\extd\) on \(\frA\) rather than \(\tilde\frA\). Projecting to \(\frA\) may increase the order of the operator, and will require some explicit verification.
Explicit verification of the order of \(\theta\extd\star_3\extd\) will require us to consider the following non-trivial arguments of the Koszul bracket: \((\vec c,B^+)\), \((\vec c,A_1,A_2)\), \((\vec c,A,B^+)\), \((\vec c,B^+_1,B^+_2)\), \((\vec c,A_1,A_2,A_3)\), \((\vec c,A_1,A_2,B^+)\) and \((\vec c,A_1,A_2,A_3,A_4)\).
The leading contribution to the order of \(\theta\extd\star_3\extd\) will come from the argument \((\vec c,A_1,A_2,A_3,A_4)\). Applying the higher-order Leibniz rule \eqref{eq:higher-order leibniz identity} gives us
\begin{equation}
\label{eq:computation Koszul brackets}
\begin{split}
	&\Phi^{r+4}_{\theta\extd\star_3\extd}\Big(\pi(\vec c),\pi(A_1),\pi(A_2),\pi(A_3),\pi(A_4)\Big)
	\\
	&= \frac1{4!}\sum_{\sigma\in S_4}\varepsilon(\sigma)\pi\Big(6\theta\Phi^{r+1}_{\extd\star_3\extd}\big(\vec c,A_{\sigma(1)}\wedge A_{\sigma(2)}\big)\wedge A_{\sigma(3)}\wedge A_{\sigma(4)}\Big)
\end{split}
\end{equation}
Following from the fact that \(\extd\star_3\extd\) is of second order in covariant derivatives and also of second order in interior derivatives, it follows that \eqref{eq:computation Koszul brackets} vanishes for \(r\geq 3\). From this it follows that the order of \(\theta\extd\star_3\extd\) on \(\frA\) is six,
\begin{equation}
\label{eq:order third term}
	\Order_\frA\big(\theta\extd\star_3\extd\big) = 6\,.
\end{equation}

Combining the results from \eqref{eq:order first term}, \eqref{eq:order second term} and \eqref{eq:order third term}, we conclude following \eqref{eq:order tilde sfb as maximum} that the order of \(\tilde\sfb\) is six,
\begin{equation}
\label{eq:order of tilde sfb}
	\Order_\frA\big(\tilde\sfb\big)
	=
	6\,.
\end{equation}

\paragraph{Order of \([\sfQ,\sfw]\).} To compute the order of the \([\sfQ,\sfw]\) term of \(\sfb\), we note that
\begin{equation}
\label{eq:Qw order inequality}
	\Order_\frA\big([\sfQ,\sfw]\big) 
	\leq \Order_\frA\big(\sfw\big) + \Order_\frA\big(\sfQ\big) - 1
	= \Order_\frA\big(\sfw\big)\,.
\end{equation}
Thus, from the order of \(\sfw\) we can get an upper bound on the order of \(\sfb\), which, if it is less than the order of \(\tilde\sfb\), will give us the order of \(\sfb\). Making use of \eqref{eq:brace degrees} and the fact that \(\frA\) is concentrated in degrees \(0,1,2,3\), we find that the non-trivial Koszul brackets are constrained to arguments \(\vec x\) obeying degree restrictions
\begin{equation}
	2 \leq |x_1| + \cdots + |x_k| \leq 5\,.
\end{equation}
The leading contribution to the order of \(\sfw\) will then come from the argument \(\vec x = (\vec c,A_1,A_2,A_3,A_4,B\theta)\), for which application of the higher-order Leibniz rule \eqref{eq:higher-order leibniz identity} yields
\begin{equation}
\begin{split}
	&\Phi^{r+5}_\sfw\Big(\pi(\vec c),\pi(A_1),\pi(A_2),\pi(A_3),\pi(A_4),\pi(B\theta)\Big)
	\\
	&= \frac1{4!}\sum_{\sigma\in S_4}
	\begin{aligned}[t]
		&-6\varepsilon(\sigma)\Phi^{r+1}_\sfw\Big(\pi(\vec c),\pi\big(A_{\sigma(1)}\wedge A_{\sigma(2)}\big)\Big)\wedge\pi\big(A_{\sigma(3)}\wedge A_{\sigma(4)}\wedge B\theta\big)
		\\
		&+ 6\varepsilon(\sigma)\pi\big(A_{\sigma(1)}\wedge A_{\sigma(2)}\big) \wedge \Phi^{r+1}_\sfw\Big(\pi(\vec c),\pi\big(A_{\sigma(3)}\wedge A_{\sigma(4)}\wedge B\theta\big)\Big)
	\end{aligned}
	\\
	&= \frac1{4!}\sum_{\sigma\in S_4}6\varepsilon(\sigma)\pi\Big(
	\begin{aligned}[t]
		&-\Phi^{r+1}_{K\intprod}\big(\vec c,A_{\sigma(1)}\wedge A_{\sigma(2)}\big)\wedge A_{\sigma(3)}\wedge A_{\sigma(4)}\wedge B\theta
		\\
		&\mp A_{\sigma(1)}\wedge A_{\sigma(2)} \wedge \Phi^{r+1}_{K\intprod}\big(\vec c,A_{\sigma(3)}\wedge A_{\sigma(4)}\wedge B\big)\theta\Big)
	\end{aligned}
\end{split}
\end{equation}
which vanishes for \(r\geq 1\). Accordingly, the order of \(\sfw\) is six, and therefore following \eqref{eq:Qw order inequality}, the order of \([\sfQ,\sfw]\) is at most six,
\begin{equation}
\label{eq:order of Qw upper bound}
	\Order_\frA\big([\sfQ,\sfw]\big) \leq \Order_\frA\big(\sfw\big) = 6\,.
\end{equation}
\paragraph{Order of \(\sfb\).} Recall that following \eqref{eq:order of tilde sfb}, the order of \(\tilde\sfb\) is six, and that the order of \([\sfQ,\sfw]\) is at most six. From this we conclude that the order of \(\sfb = \tilde\sfb + [\sfQ,\sfw]\) is also six,
\begin{equation}
	\Order_\frA\big(\sfb\big) = 6\,.
\end{equation}
This is because terms arising from the deformation \([\sfQ,\sfw]\) always contain \(K\), which cannot cancel terms coming from \(\tilde\sfb\).

    \end{body}
\end{document}